\documentclass[onecolumn,groupedaddress]{revtex4}
\usepackage{graphicx}
\usepackage{epstopdf}
\usepackage{latexsym}
\usepackage{amssymb}
\usepackage{amsfonts} 
\usepackage{txfonts}
\usepackage{pxfonts}
\usepackage{color}
\usepackage[usenames,dvipsnames]{xcolor}
\usepackage{pdfpages}

\newcommand{\kstiff}{\ensuremath{k_{\theta}}}
\newcommand{\dd}{\textrm{d}}

\begin{document}

\title{The role of shape disorder in the collective behaviour of aligned fibrous matter}
\author{Salvatore Salamone, Nava Schulmann, Olivier Benzerara, Hendrik Meyer,  Thierry Charitat$^{\ast }$, Carlos M. Marques}
\affiliation{Institut Charles Sadron, Universit\'e de Strasbourg, CNRS, 23 rue du Loess, BP 84047 67034 Strasbourg Cedex 2, France}
\date{\today}

\begin{abstract}
We study the compression of bundles of aligned macroscopic fibers with intrinsic shape disorder, as found in human hair and in many other natural and man-made systems. We show by a combination of experiments, numerical simulations and theory how the statistical properties of the shapes of the fibers control the collective mechanical behaviour of the bundles. 
This work paves the way for designing aligned fibrous matter with pre-required properties from large numbers of individual strands of selected geometry and rigidity.
\end{abstract}

\maketitle

\section{Introduction}

In natural and composite bundles of nearly fully aligned fibers, as for instance in hair tresses, ponytails and other natural fagots, the spontaneously curved shapes of the individual strands allow for an intrinsic fluffiness of the materials~\cite{2002_robbins,2010_audoly_pomeau,1999_baudequin_roux,2007_kabla_mahadevan}. This was first discussed by Van Wyk \cite{1946_van-wyk} who proposed an equation of state (EOS) for the material compressibility by suggesting that the response of wool stacks to compression is mostly controlled by the bending modes of the fiber strands. The suggestion has been widely discussed in work related to fibrous matter~\cite{1987_gutowski_wineman}, in explanations of the compressibility of textiles~\cite{1993_pan}, matted fibers~\cite{2005_poquillon_andrieu}, non-woven fibrous mats, needled \cite{soares(2017)} or not \cite{siberstein(2012)}, of bulk samples of wool fibers~\cite{1955_demacarty_dusenbury}, for studies of the shape of hair ponytails~\cite{2012_goldstein_ball}, for predicting droplet formation at the tip of wet brushes~\cite{2016_yamamoto_doi}, or even to study  the mechanical response of aegagropilae~\cite{2017_verhille_le-gal}. 

Since van Wyk's seminal work \cite{1946_van-wyk} many models have been proposed to understand the collective mechanical elasticity of stacks of randomly oriented straight fibers \cite{wilhelm2003elasticity,gardel2004elastic,buxton2007bending,broedersz2011criticality}. Recently Broedersz {\it et al.} have shown that the elastic properties of such networks are governed by bending elasticity for low connectivity and by stretching elasticity for high connectivity \cite{broedersz2011criticality}.
\begin{figure}[h!]\begin{center}
\includegraphics[width=0.5\textwidth]{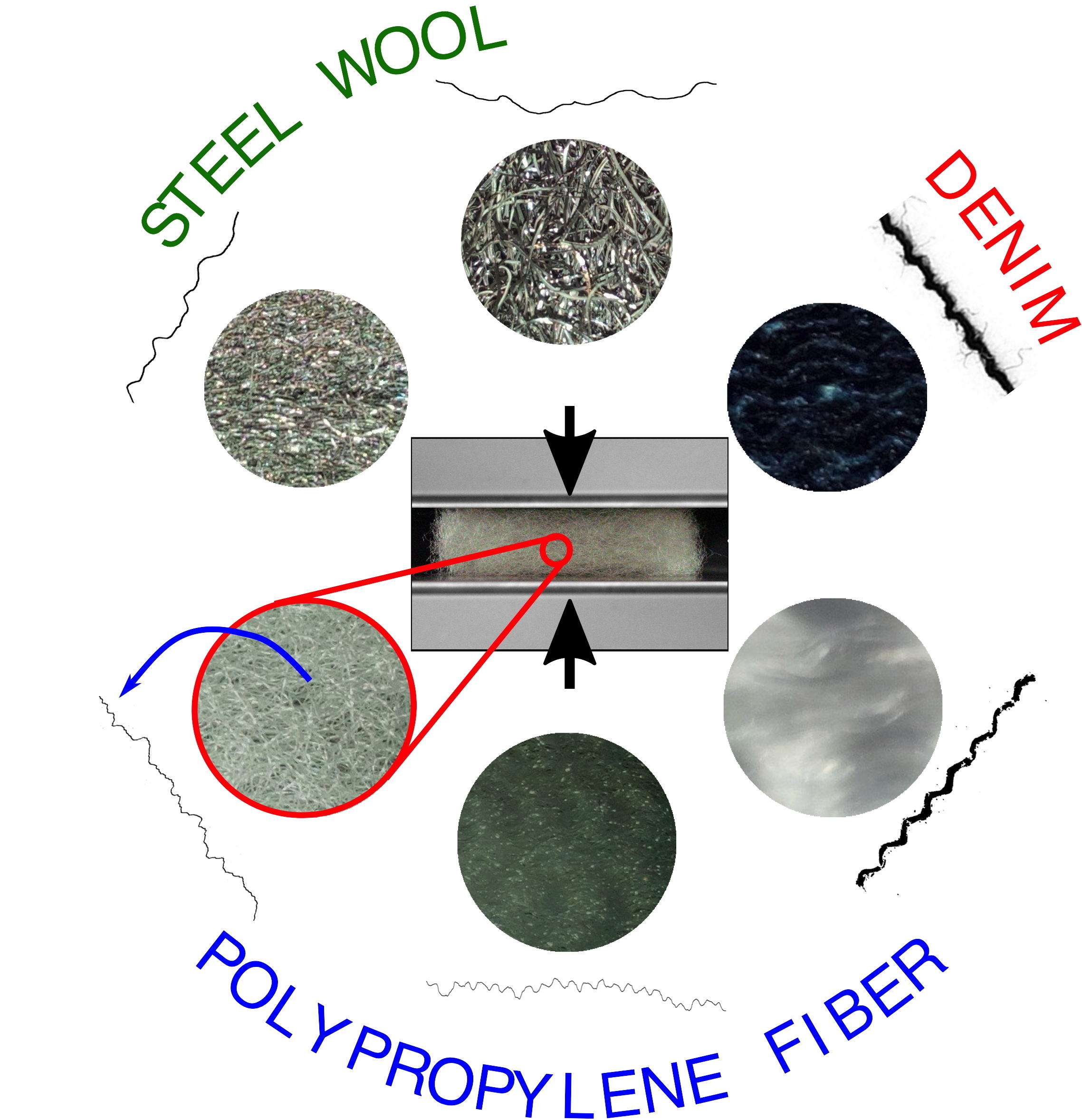}
\caption{Central image: compression of a polypropylene stack and summary of the different samples (steel wool, denim and polyprolpylene fiber) investigated in this paper. Circular vignettes: zoom on a volume of typical size 1 cm. Surrounding pictures: typical shapes of individual fibers of each sample ($\sim 10$ cm length).}
\label{figure1}
\end{center}\end{figure}
In this paper, we focus on the case of highly aligned fibers with disordered shapes. The statistical mechanics nature of this challenge was first recognised by Beckrich {\sl et al.}~\cite{2003_beckrich_charitat} who computed the compression modulus of fiber stacks within a self-consistent mean-field treatment in two dimensions, predicting the shapes of brooms and other fluffy cones made from fibers.  Here we test the validity of a statistical mechanics treatment of this problem by studying both experimentally and numerically the compression behaviour of stacks of fibers with intrinsic shape disorder (see Fig.\ref{figure1}). A generalisation of the mean field theory introduced in \cite{2003_beckrich_charitat} compares well with our results, revealing the key statistical and mechanical factors that control the EOS of fibrous matter.

\section{Materials and methods}

\subsection{Experimental systems}

\begin{figure}[h!]
\begin{center}
\includegraphics[width = 0.4\textwidth]{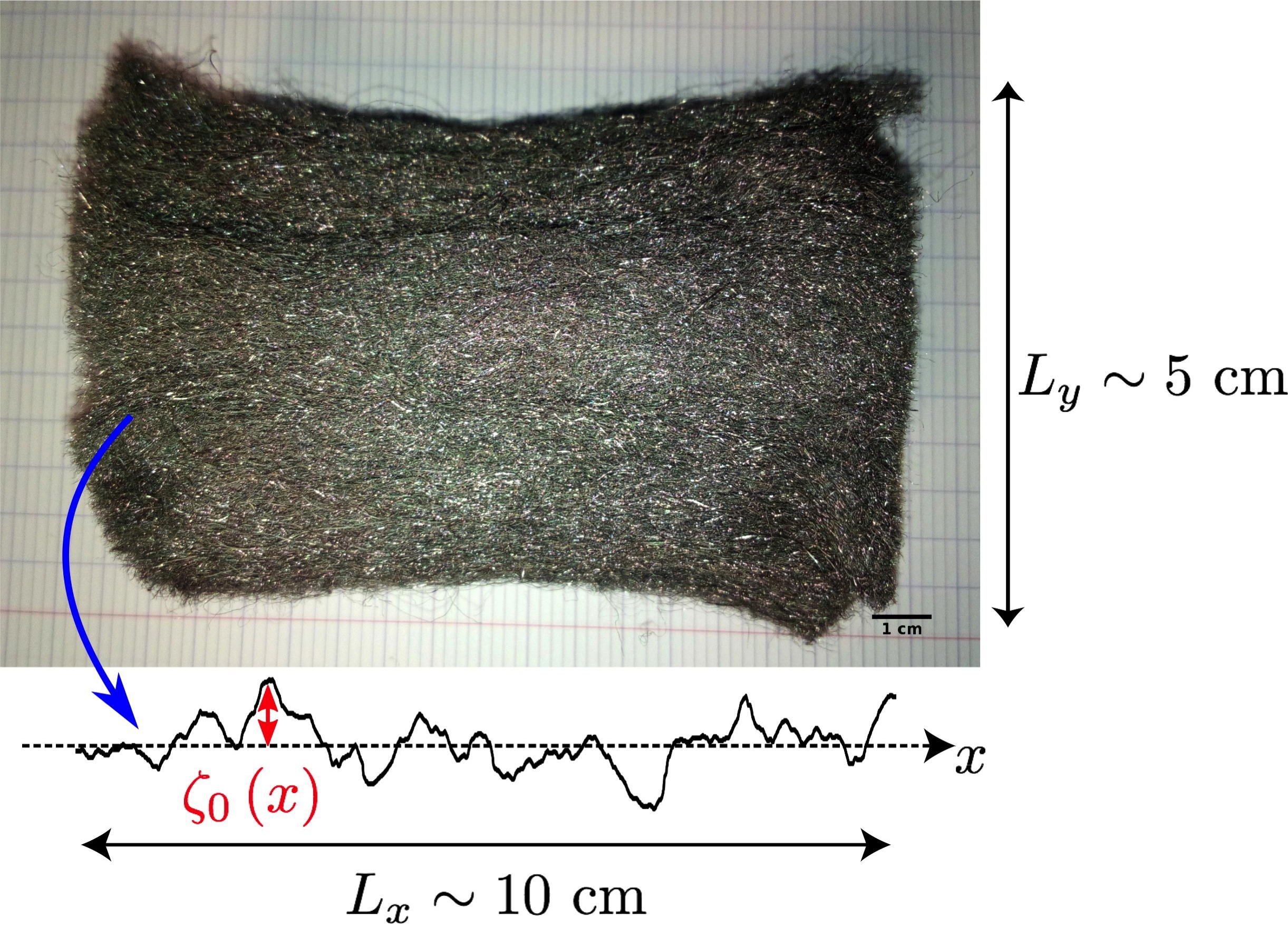}
\caption{(Top) Example of a fiber stack of steel wool (SW1) with typical characteristic length scales ($L_x, L_y$). (Bottom) Single fiber shape and definition of $\zeta_0\left(x\right)$.}
\label{figure2}
\end{center}
\end{figure}

Experiments were performed on 6 different fiber bundles with caracteristic sizes noted $L_x, L_y$ and $D_0$, see Fig. \ref{figure2} and \ref{figure3}. A first class of samples was obtained by unbraiding different commercial climbing ropes of polypropylene (PP1, PP2 and PP3) and standard denim cloth (DEN). For most fiber studies, fiber bundles were formed by stacking manually a high number (hundreds or thousands) of individual fibers of the same length, ensuring a strong alignment.
A second class of samples (SW1 and SW2) was obtained from standard steel wools (Gerlon\textregistered). In this case, compression experiments were directly performed on the purchased samples.

The fiber sections, as observed by optical microscopy, do not have regular shapes, and we measured the following approximate diameters: PP1 and PP2 ($\sim 100\pm 20\ \mathrm{\mu}$m); PP3 ($\sim 20\pm 5\ \mathrm{\mu}$m); DEN ($\sim 500\pm 100\ \mathrm{\mu}$m); SW1 ($\sim 200\pm 50\ \mathrm{\mu}$m) and SW2 ($\sim 300\pm 100\ \mathrm{\mu}$m) (see ESI section S1). 

The individual mass of several fibers of each stack was measured allowing us to determine the mass per length $\mu$. Together with the full mass of the stack, it allows us to estimate the number $N$ of fibers per bundle and the transverse density $\rho=\sqrt{N}/L_y$ (see Table \ref{stretch-table-1}). 

\subsection{Experimental methods}

\begin{figure}[h!]
\begin{center}
\includegraphics[width = 0.4\textwidth]{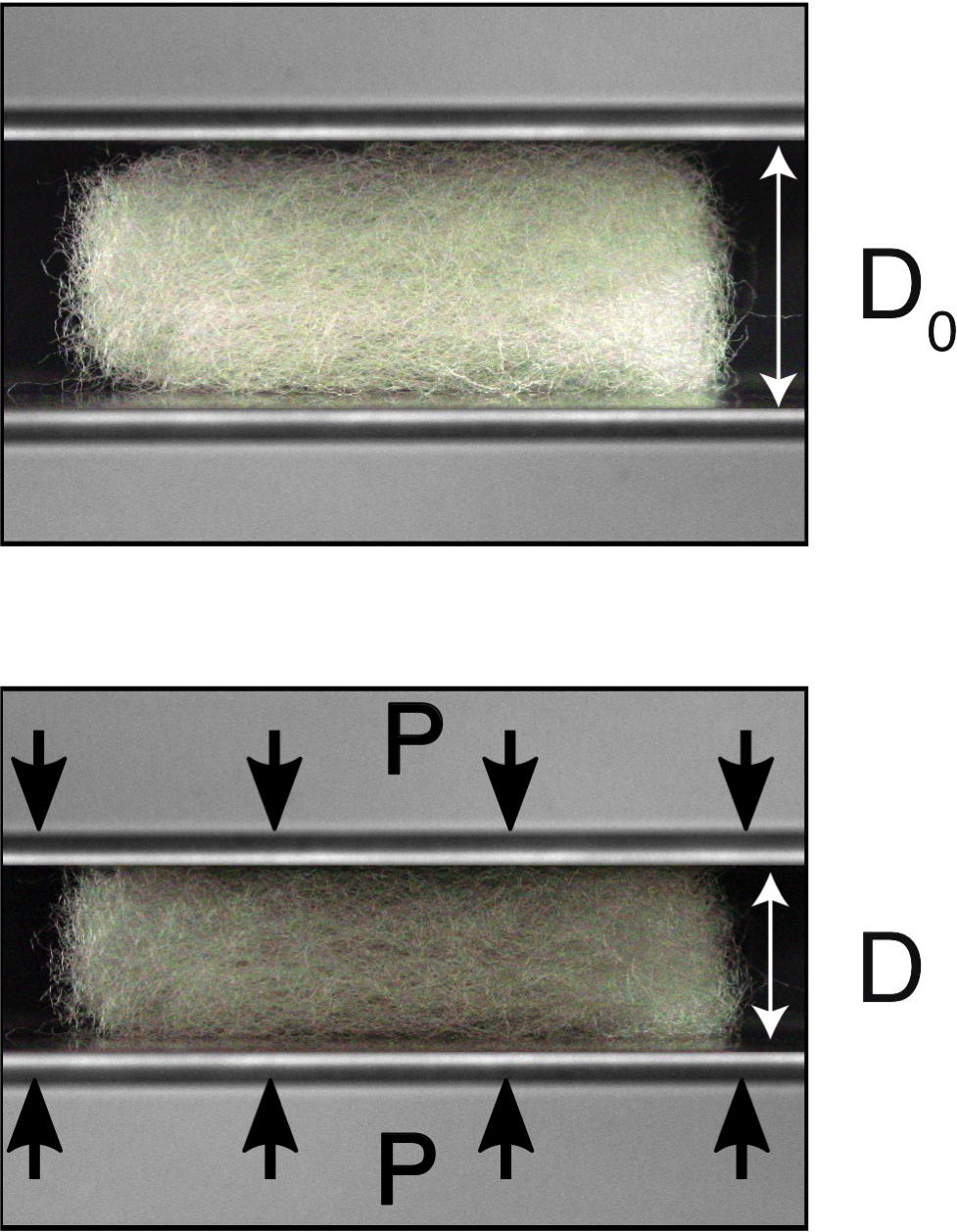}
\caption{Experimental images of a stack of thousands of fibers (PP1) before compression (top) and under a compression stress $P$ (bottom) and definitions of $D$ and $D_0$.}
\label{figure3}
\end{center}
\end{figure}

Compression experiments were made using two different experimental setups. 

The first one is homemade, from a precision balance used as a force sensor (Mettler Toledo\textregistered) and a motorized translation stage. This setup provides a very good resolution (accuracy, $\pm\ 2\times 10^{-4}$ N) in a wide range of force (6 decades from $2\times 10^{-4}$ N to 40 N). The stiffness of the device is of the order of 10 MPa, and we have systematically corrected for scale plate displacement.

The second setup is a commercial testing system (Electropuls\texttrademark E3000 from Instron\textregistered, 10 N sensor), with a lower force range (3 decades $10^{-2}$-10 N, accuracy $\pm\ 10^{-2}$ N) but a higher stiffness ($\sim$ 100 MPa). This setup was mainly used for stress-relaxation experiments.

With both setups and for each experiment we measured the distance $D_0$ between the compression plates at first contact, the distance $D$ between compression plates at each step of compression, the projected lengths $L_x$ and $L_y$ (see Fig. \ref{figure2}) and the force $F$. We calculated the stress $P=F/L_xL_y$ and the relative deformation that we define as $D_0/D$ (see Fig. \ref{figure3}).

\subsection{Theoretical description}

We introduce now the theoretical models used in the following. We assume that we have a two-dimensional stack of ${\cal N}$ fibers in between two hard walls separated by a distance $D$ for a compression stress $P$ ($D=D_0$ for $P=0$). The average distance between fibers will be noted $d=D/{\cal N}$ (and $d_0=D_0/{\cal N}$ for $P=0$). Individual fiber shape deformations are associated to the bending energy given by:
\begin{equation}
{\cal H}_{\rm bend}=\kappa /2 \sum_{n=1}^{\cal N} \int_0^{L_x} \textrm{d}x\left[\zeta_n ''(x)- \zeta''_{0,n}(x) \right]^2
\label{Hbending}
\end{equation}
where $\zeta_n(x)$ is the deformed shape and $\zeta_{0,n}(x)$ the function describing the spontaneous shape of fiber $n$ (see Fig. \ref{figure2}). $L_x$ corresponds to the chain projection on the $x$ axis. The bending modulus $\kappa$ is an intrinsic property of a fiber, directly related to the Young's modulus and the fiber geometry. Equation (\ref{Hbending}) is valid in the limit of small deformation gradients ($\zeta_n^\prime\ll 1$) that is a very good approximation in the case of nearly aligned fiber stacks as we checked on experimental and numerical systems.

\subsection{Numerical simulations}

\begin{figure}[h!]
\begin{center}
\includegraphics[width = 0.4\textwidth]{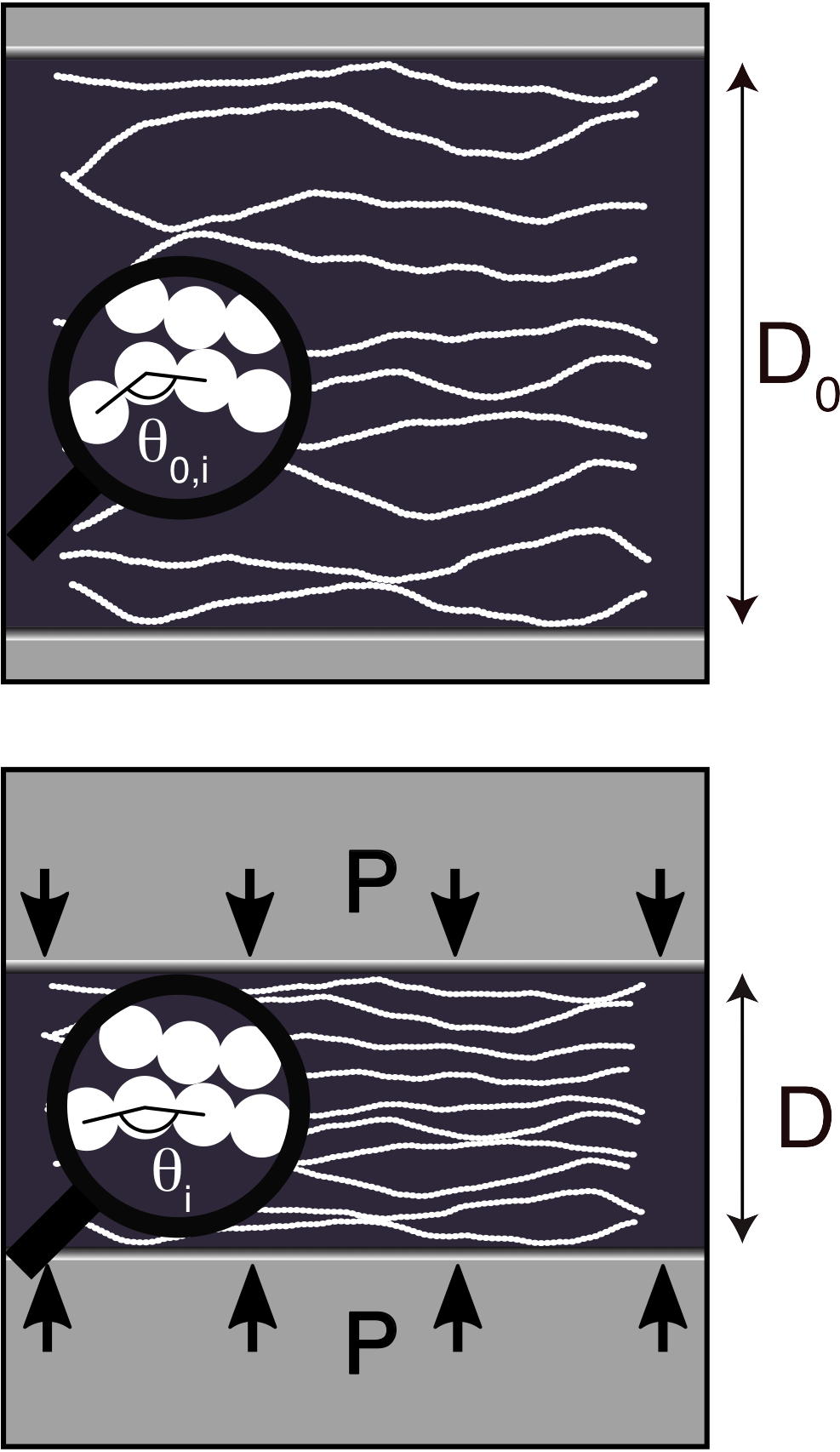}
\caption{Sketch of the  geometry of the numerical simulations at two different compression distances. The insets show the shapes of two neighbouring fibers with the definition of the angles $\theta_{0,i}$ and $\theta_{i}$ between consecutive beads $i-1$, $i$ and $i+1$.}
\label{def-simu}
\end{center}
\end{figure}

Numerical simulations were performed using the steepest descent method \cite{1952_hestenes_stiefel} to find the equilibrium conformations of compressed fibers represented by the bead-spring model \cite{1986_grest_kremer,1990_kremer_grest} sketched in Fig.~\ref{def-simu} with $N_b$ beads per fiber. Interactions between beads are modelled by an effective Hamiltonian containing three terms: 
\begin{equation}
{\cal H}={\cal H}_{\rm LJ}+{\cal H}_{\rm bond}+{\cal H}_{\rm angle}.
\label{hamiltoniennum}
\end{equation}
The first term corresponds to the truncated and shifted Lennard-Jones (LJ) potential \cite{2001_frenkel_smit} describing the repulsive interaction between non-neighboring beads 
\begin{equation}
{\cal H}_{\rm LJ}=4\epsilon\left[(\sigma/r)^{12}-(\sigma/r)^{6}\right]+ \epsilon \ \ \mbox{for}\ \ r /\sigma \le 2^{1/6},
\label{eq_algoLJ}
\end{equation} 
where $r_{i,j}$ is the distance between two monomers $i$ and $j$, and $\sigma$ the monomer size. The second term is the connectivity potential between two adjacent monomers on the same fiber
\begin{equation}
{\cal H}_{\rm bond} = \frac{k_b}{2}\ (r_{i,i+1} -r_{0})^2,
\label{H_connect}
\end{equation} 
with a strong spring constant $k_b=600\epsilon$ and $r_0=\sigma$ the distance between two connected beads for a non-deformed fiber. The last term is the discrete representation of eqn (\ref{Hbending}). It corresponds to the angular potential that controls the chain stiffness and spontaneous shape,
\begin{equation}
{\cal H}_{\rm angle} = \frac{\kstiff}{2}\ (\theta_i -\theta_{0,i})^2,
\label{H_anglet}
\end{equation} 
with $\theta_i$ the angle between bonds $(i-1,i)$ and $(i,i+1)$ and $\kstiff$ the angular stiffness. The set of non-vanishing reference angles $\{\theta_{0,i}\} \left(i=1,...,N_b-1\right)$ between any three consecutive monomers -- see Fig.~\ref{def-simu} -- are chosen such that the local fiber gradients remain much smaller than unity. Fiber shapes can thus be also described by a single-valued function $\zeta_0\left(x\right)$ which allows, in the limit of large fibers, to directly compare numerical results against continuous elastic theories with $\kappa=\kstiff r_0$ the bending modulus.

\section{Single fiber characterization}

\subsection{Shape characterization}

\begin{figure}[h!]\begin{center}
\includegraphics[width=0.5\textwidth]{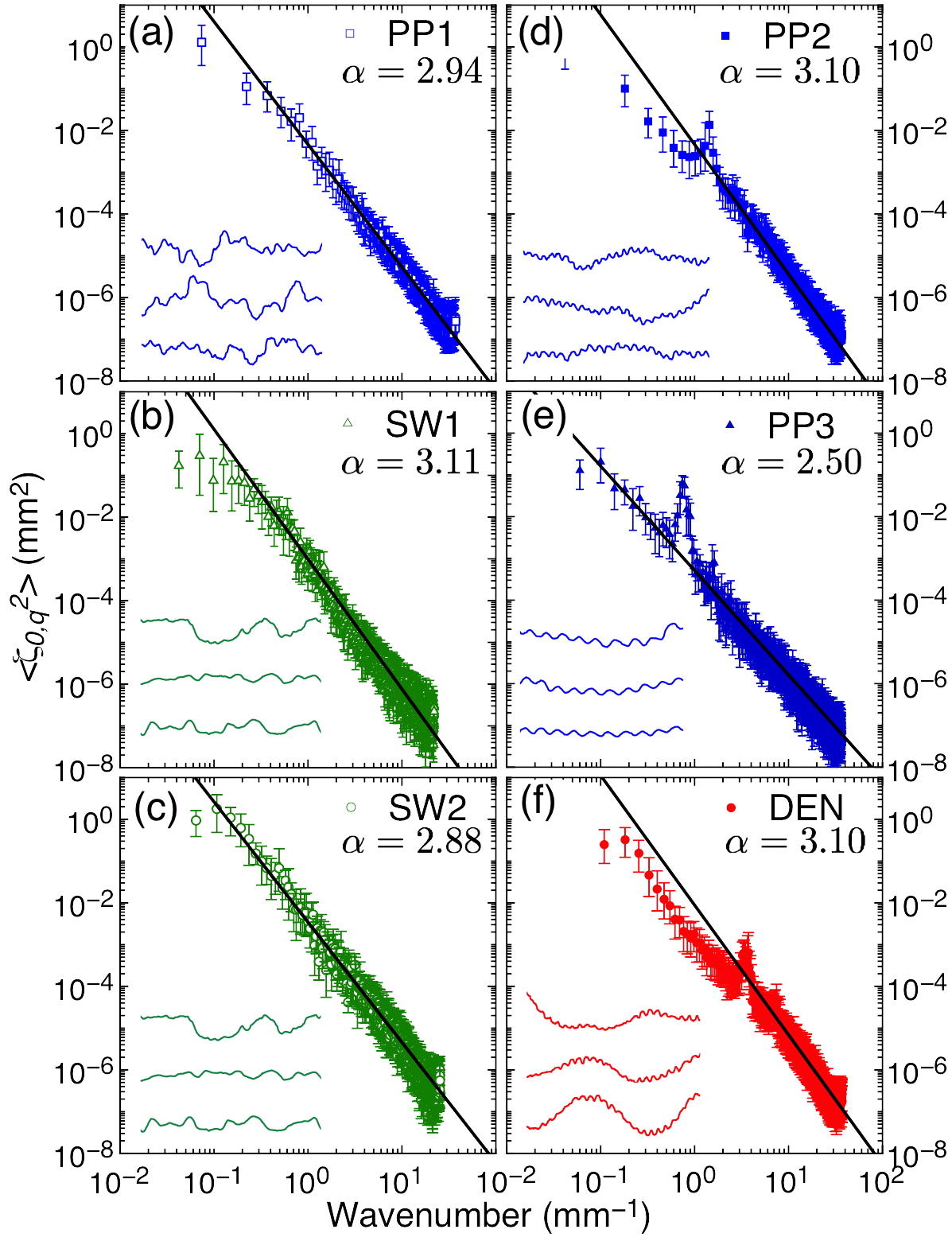}
\caption{(Color online) Average spectra $\langle\zeta_{0,q}^2\rangle$ extracted from measurements of at least 40 fibers of the 6 experimental systems, and best power law fits with exponent $\alpha$ (solid line) for PP1 (a), SW1 (b), SW2 (c), PP2 (d), PP3 (e) and DEN (f). The insets display typical experimental fiber shapes.}
\label{figspectre}
\end{center}\end{figure}

As we will see below, the spontaneous shape of the fibers $\zeta_0(x)$ is a crucial determinant of the collective mechanical behavior of the bundle. We measure  $\zeta_0(x)$ for a large number ($\sim 50$) of individual fibers and we expand $\zeta_0(x)$ on the basis of the eigenfunctions  $\Phi_{q_i}(x)$ of the square Laplacian operator that describes bending curvature elasticity \cite{love(1892),1986_landau_lifshitz} (see also the ESI section S1 for more details), $\zeta_0(x) = \sum_i  \zeta_{0,q_i} \Phi_{q_i}(x)$. The corresponding spectra for average values $\langle \zeta_{0,q}^2\rangle$ are displayed on Fig.~\ref{figspectre}. The dominant amplitudes are present for $q \leq 2$ mm$^{-1}$. The dispersion of the variance spreads over 6 decades. A dominant wavelength appears very clearly for PP2, PP3 and DEN. All spectra exhibit a power law regime over two or more decades of wavenumber values. The power exponent denoted $\alpha$ is related to the roughness of the fiber and the prefactor of the power-law to its root mean square amplitude.

\subsection{Mechanical properties}

To determine experimentally the bending modulus $\kappa$ of fibers we performed two different types of experiments. The first one consists in measuring the oscillation period of a horizontal fiber (see section \ref{sectionoscillation}). This method is particularly suited for steel fibers (SW1 and SW2) which are sufficiently rigid. However, we have not been able to apply it to synthetic fibers (PP1, PP2, PP3 and DEN) which are too light and too sensitive to small disturbances. For these fibers we have developed an original experience of stretching, inspired by the work of Kabla and Mahadevan \cite{2007_kabla_mahadevan} (see section \ref{sectionfolding}).

\subsubsection{Single fiber oscillations}
\label{sectionoscillation}

\begin{figure}[h!]
\begin{center}
\includegraphics[width = 0.4\textwidth]{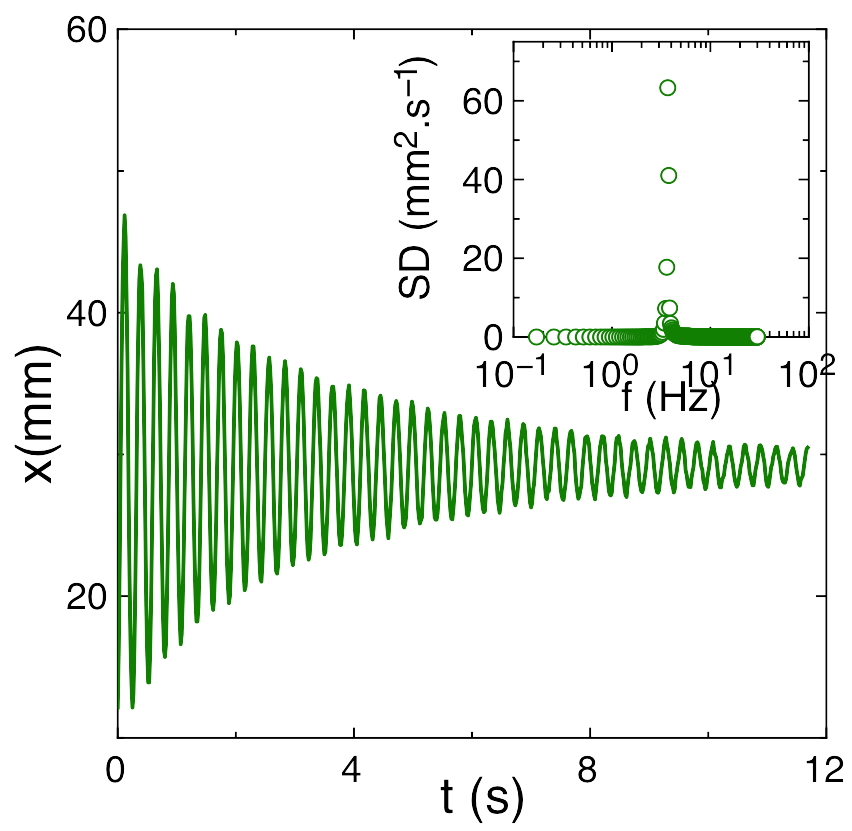}
\caption{Position of the end point of a SW2 fiber during an oscillation experiment. In the inset, the spectral density (SD) obtained by a simple Fourier Transform of the signal.}
\label{oscillations}
\end{center}\end{figure}

The position of the end of a horizontal fiber is measured over time during an oscillation experiment (see Fig. \ref{oscillations}, see also the ESI for a video). A simple Fourier transform allows precisely characterising the fundamental frequency $f_0$ of the system (see Fig. \ref{oscillations} inset), which is related to the modulus of curvature by the equation \cite{love(1892),1986_landau_lifshitz}:
\begin{equation}
\kappa = 3.194 \times m L^4 f^2_0,
\end{equation}
where $m$ is the fiber mass and $L$ its length.

\subsubsection{Single fiber stretching experiments}
\label{sectionfolding}

To determine experimentally the bending modulus $\kappa$ of softer fibers we performed stretching experiments using the same experimental set-up as for fiber stack compression. A single fiber is attached by its extremities to the scale plate and the translation stage (see the ESI for a video). At each stage of the experiment, we measure the stretching force $f$ and a snapshot is taken whereby the total length, the projected length and the shape analysis of the fiber are calculated. To avoid creating new folding states during the preparation of the fiber, we proceed as follows. 
We first measure the projection length $L_0$ of the fiber along the average direction of the unperturbed free fiber. 
We then fix the fiber extremities to the experimental set-up, and impose a distance between extremities slightly smaller than 
$L_0$. Forces and distances are then only considered for distances equal or above to $L_0$.

\begin{figure}[h!]
\begin{center}
\includegraphics[width = 0.4\textwidth]{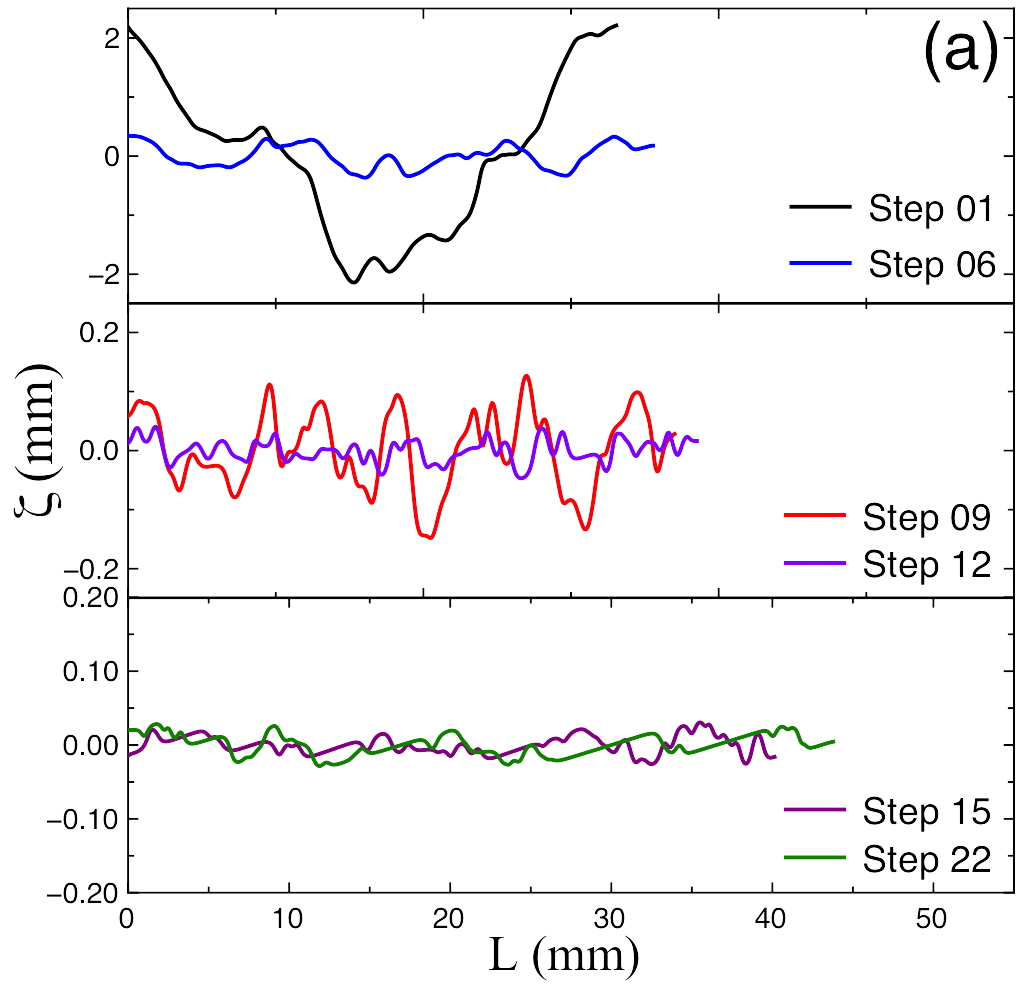}
\caption{Shape of a PP1 fiber at different steps of a stretching experiment.}
\label{stretch-1}
\end{center}
\end{figure}

As an example, the shape of a PP1 fiber for a few steps of a stretching experiment is shown in Fig. \ref{stretch-1}.

In this example, the reference state ($f\left(L=L_0\right)=0$) corresponds to step 6. Steps 6 to 12 show clearly that the largest wavelengths are first unfolded, as confirmed by the spectra (see the ESI section 1), where the amplitudes of the modes $q_1 \simeq 0.5$ mm$^{- 1}$ and $q_2 \simeq 1.0$ mm$^{- 1}$ (ie $\lambda_1=2\pi/q_1 \simeq$ 12 mm and $\lambda_2= 2\pi/q_2 \simeq$ 6 mm) decreased significantly more than the other ones. Beyond step 15, the deformation is dominated by elongation. Complete mode supression is typically not attainable, the fibers breaking before becoming completely straight.

\begin{figure}[h!]\begin{center}
\includegraphics[width = 0.4\textwidth]{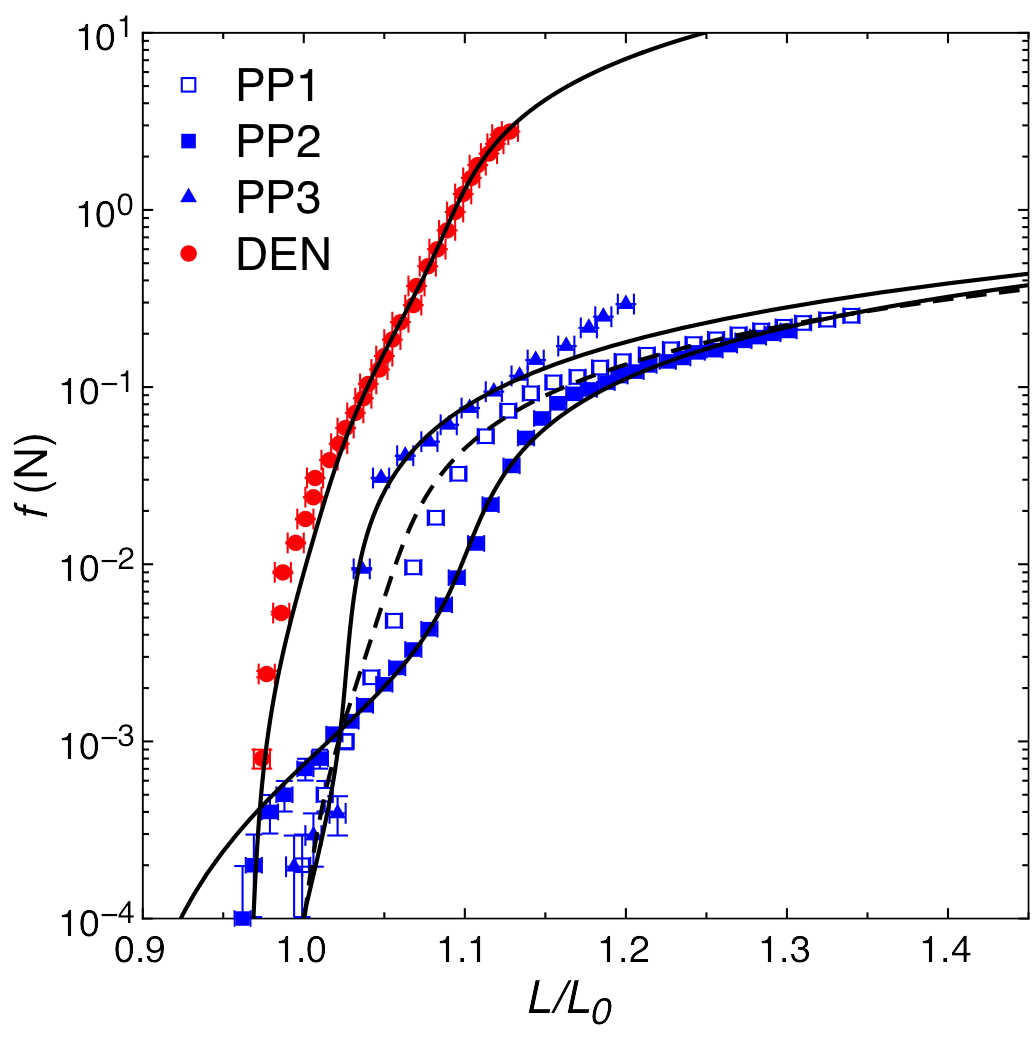}
\caption{Measured force as a function of the ratio $L/L_0$ for PP1, PP2, PP3 and DEN samples. Solid lines corresponds to best fits obtained using eqn (\ref{equation-stretching-1}).}
\label{etirement-2}
\end{center}\end{figure}

The force measured during the stretching experiment is shown in Fig. \ref{etirement-2} as a function of the length ratio. Using the equation of state developped for an extensible fiber by Kabla and Mahadevan \cite{2007_kabla_mahadevan}
\begin{equation}
\frac{L}{L_0}=1+\frac{f}{\mu_0} -\frac14(1+2\frac{f}{\mu_0})\sum_q\frac{q^6\zeta_{0,q}^2}{(f/\kappa+q^2)^2}+\frac{\kappa}{2\mu_0}\sum_q\frac{q^6\zeta_{0,q}^2}{f/\kappa+q^2},
\label{equation-stretching-1}
\end{equation}
it is possible to deduce the bending modulus $\kappa$ (J.m) and the stretching modulus of the fiber $\mu_0$ (N).

Best fits of the experimental curves by eqn (\ref{equation-stretching-1}) are shown in Fig. \ref{etirement-2} and the moduli are summarized in Table \ref{stretch-table-1}.

\subsection{Single fiber characteristics}

As a summary, we carefully characterized the shapes and the bending modulus of individual fibers of the six samples that we studied. We computed the average shape spectrum of the different samples ($\{q_i,\langle\zeta_{0,q_i}^2\rangle\}$) and the average of the bending modulus. The results are summarised in Table \ref{stretch-table-1}.

\begin{table}[h!]
\small
\caption{Mean value of linear mass density ($\mu$), bending ($\kappa$) and stretching ($\mu_0$) moduli and transerve density ($\rho$).}
\label{stretch-table-1}
\begin{center}
  \begin{tabular}{lclclclclclcl}
  \hline
 & $\mu$ ($\rm{\mu}$g.cm$^{-1}$ ) &$\kappa \times10^{12}$ (J.m) & $\mu_0$ (N) & $\rho$ (mm$^{-1}$)
 \\
\hline
\hline
PP1& 24 $\pm$ 1 & $3.2\pm1.7$&$1.0\pm 0.2$ & $1.2 \pm 1$
\\
\hline
PP2& 18 $\pm$ 1 & $ 3.5\pm1.8 $ & $1.3\pm 0.2$ & $2.5 \pm 2$
\\
\hline
PP3& 4.7 $\pm$ 0.5 & $0.43 \pm 0.3$ & $ 1\pm 0.2$ & $6.8 \pm 4$
\\
\hline
SW1& 110 $\pm$ 2 & $ 78 \pm 42$ & - & $ 1.6 \pm 1$
\\
\hline
SW2& 610 $\pm$ 11 & $ 160 \pm 70$ & - & $0.6 \pm 0.4$
\\
\hline
DEN& 850 $\pm$ 10 & $ 3800 \pm 3200$ & 92$\pm$ 46 & $ 1 \pm 0.5$
\\
\hline
\end{tabular}
\end{center}
\end{table}

\section{Mechanical properties of fiber stacks}

\subsection{Theoretical description}

\subsubsection{Ordered stacks: linear elasticity}

A perfectly ordered system, consisting of a stack of sinusoidal fibers $\zeta_{0,i}(x)= d_0/2 \cos(2\pi x/\lambda_i+\varphi_i)$, with ($\lambda_i=\lambda$ for all $i$, $\varphi_i=(1-(-1)^i)\pi/2$), will be referred to as the {\sl reference system} (ref) in the following. In such an ideal configuration, the EOS can be computed easily from the functional minimization of eqn (\ref{Hbending}) leading to  
\begin{equation}
P^{\rm (2d)}_{\rm ref}=192\frac{\kappa d_0}{\lambda^4}\left(\frac{D_0-D}{D_0}\right).
\label{pref}
\end{equation}
This reference model will be compared to the results of numerical simulations to validate our method.

\subsubsection{Self-consistent model}
\label{selfconsistent}

The challenge for a statistical mechanical treatment of aligned fibrous systems is thus to connect the information contained in spectra such as those of Fig.~\ref{figspectre} and the mechanical behaviour under compression stress. We follow here a two-dimensional approach first introduced in \cite{2003_beckrich_charitat}. Briefly, fiber shape deformations are associated to the bending energy given by eqn (\ref{Hbending}). We assume that forces between first-neighbours dominate the interaction energy, an exact assumption for excluded volume potentials in two-dimensions. Assuming a quadratic form for the interactions, with a compression modulus $B\left(d\right)$, the effective energy can be  written as:
\begin{equation} {\cal H}_{\rm eff}={\cal H}_{\rm bend} +
\frac{B\left(d\right)}{2} \sum_{n=1}^{\cal N}  \int_0^L
\textrm{d}x\left[\zeta_{n+1}(x)-\zeta_{n}(x)\right]^2 \ .
\label{effnrj}
\end{equation}
By functional minimization, we deduce the equilibrium shapes of the fibers and calculate the energy density:
\begin{eqnarray}
\langle e\rangle &=& \frac{\kappa}{2d}\sum_q q^4\left(1-\sqrt{\frac{q^4}{\frac{4B\left(d\right)}{\kappa}+q^4}}\right) \langle\zeta_{0,q}^2\rangle
\label{en}
\end{eqnarray}
where $d$ is the mean distance between fibers. The compression modulus $B\left(d\right)$ has to be determined self-consistently from 
\begin{eqnarray}
B_{\rm{self}}\left(d\right) = \partial ^2 \left(d\left<e\right>\right)/\partial d^2,
\label{sc}
\end{eqnarray}
the compressive stress  given by $P_{\rm{self}}^{\rm{(2d)}} = -\partial \left(d\left<e\right>\right)/\partial d$ can then be calculated using:
\begin{eqnarray}
P_{\rm{self}}^{\rm{(2d)}}(d) &= -\int_{d_0}^dB_{\rm{self}}(d')\dd d'. 
\label{p-sc}
\end{eqnarray}
Equation (\ref{en}) is the key result of this approach, it relates the shape disorder distribution measured by $\langle\zeta_{0,q}^2\rangle$ to the energy density of the stack through a mechanical kernel accounting for bending rigidity and fiber interactions and allowing to calculate self-consistently the state equation $P_{\rm self}\left(d\right)$.

\subsection{Experimental results}
 
\begin{figure}[h]
\centering
\includegraphics[angle=0,width=0.5\textwidth]{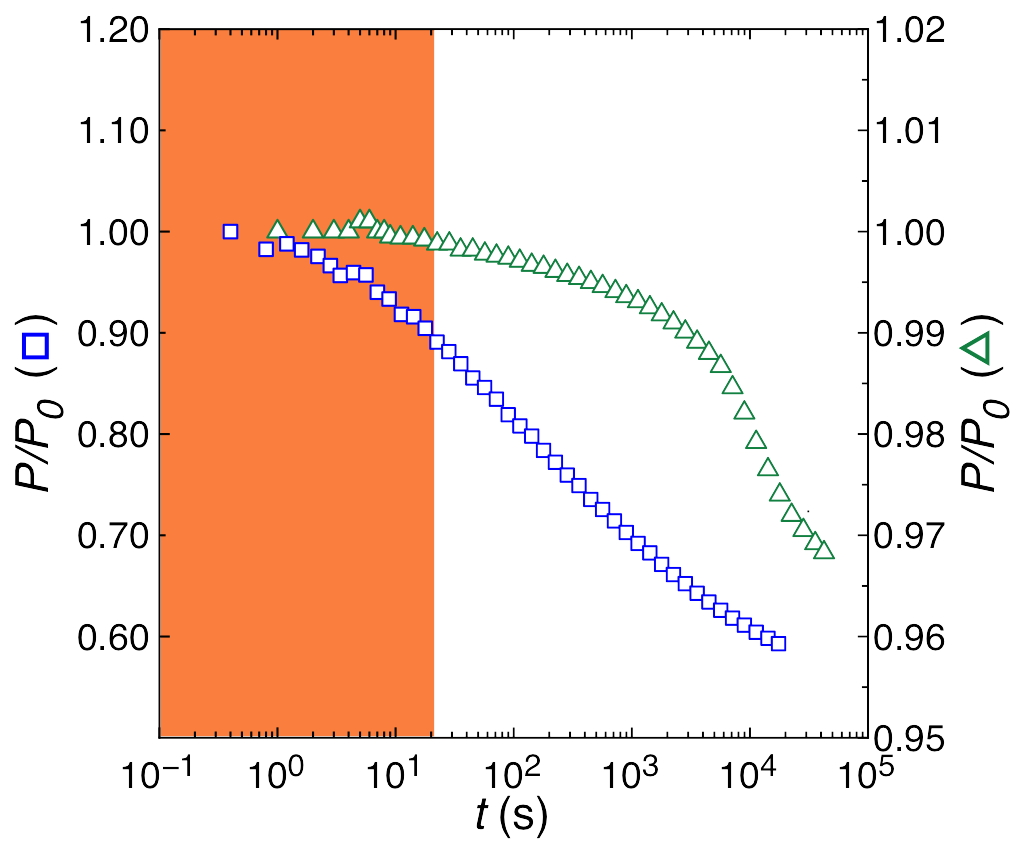}
\caption{(Color online) Normalised stress vs times  curves for PP1 ($\textcolor{blue}{\square}$, left axis) and SW1 ($\textcolor{green}{\bigtriangleup}$, right axis) at $50\%$ of deformation in 2s. The orange area corresponds to the duration of stress experiments (20 s) presented in Fig. \ref{figdefphiexp} and \ref{figpexp}.}
\label{figstressrelaxation}
 \end{figure}  

We first perform stress relaxation experiments over long times ($\sim 10$ hours). All materials demonstrate a complex relaxation behavior, highlighting the effect of fiber rearrangement (see Fig. \ref{figstressrelaxation}). Polymer samples (see PP1) exhibit a relaxation of almost 50\% over 10 hours while steel wools relaxes only about 5\% over the same time. This is likely related to the higher friction coefficient between steel fibers compared to polymers. In all cases, it is clear from the relaxation experiments that performing the full compression experiments at high enough strain rate is important in order to avoid stress relaxation by fiber rearrangement. We have thus decided to perform compression experiments on all samples with strain rate of the order of 50 mHz (full compression experiments in less than  20 s), corresponding to a maximum stress relaxation of 10\% for PP samples and less than 1\% for steel wools. 

\begin{figure}[h!]
\centering
\includegraphics[width=0.45\textwidth]{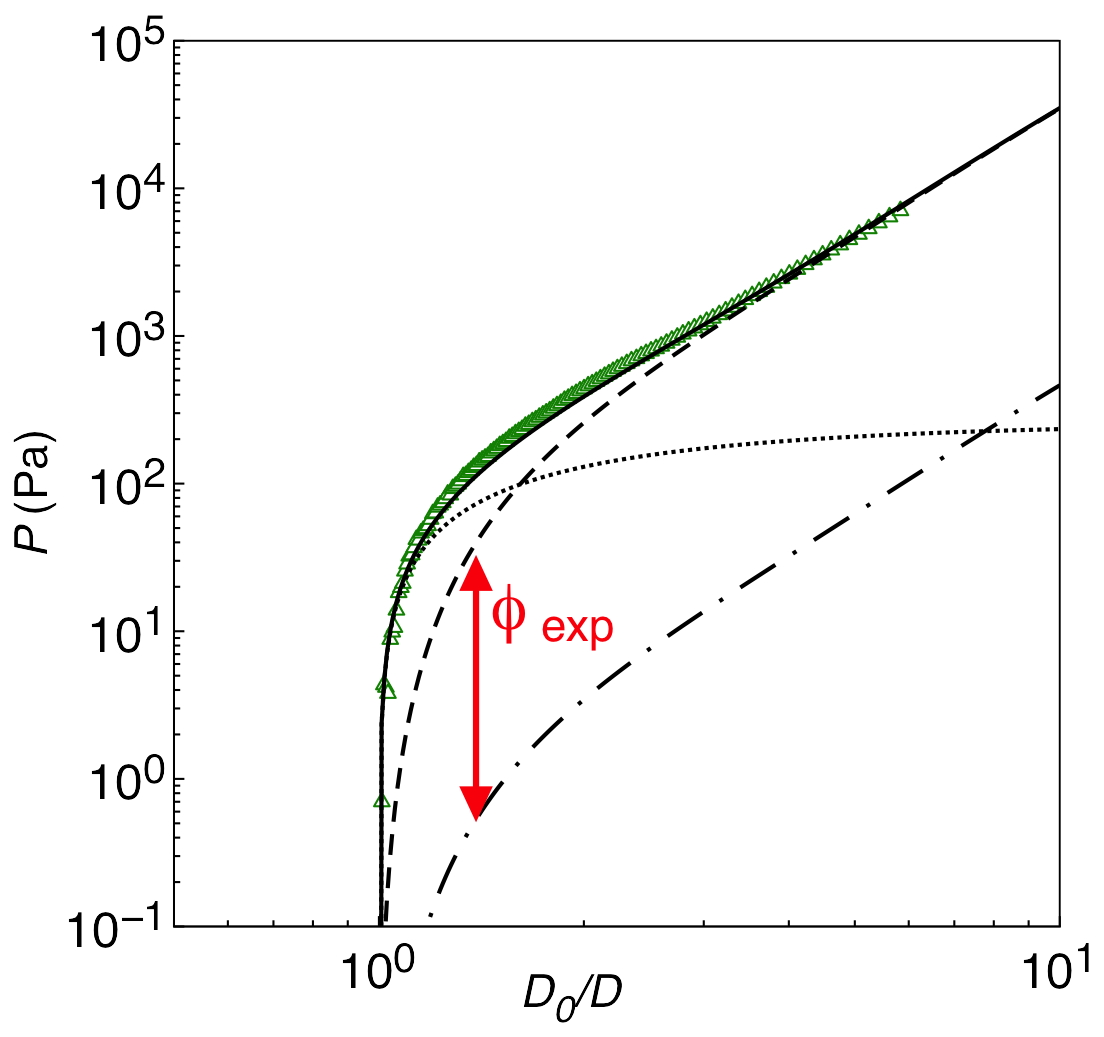}
\caption{Experimentally determined compression stress $P$ vs relative deformation $D_0/D$ for SW1 stack ($\textcolor{green}{\bigtriangleup}$). Dashed-dotted line represents the discrete self-consistent theory $\rho P^{\rm (2d)}_{\rm self}$ using the experimentally determined spectrum, the dashed line the corrected discrete self-consistent theory $\rho\Phi_{\rm exp} P^{\rm (2d)}_{\rm self}$, the dotted line the linear elasticity behavior $P_{\rm ref}$ and the solid line the total stress $P_{\rm tot}$ (eqn (\ref{ptot})).}
\label{figdefphiexp}
\end{figure}

\begin{figure}[h!]
\centering
\includegraphics[width=0.5\textwidth]{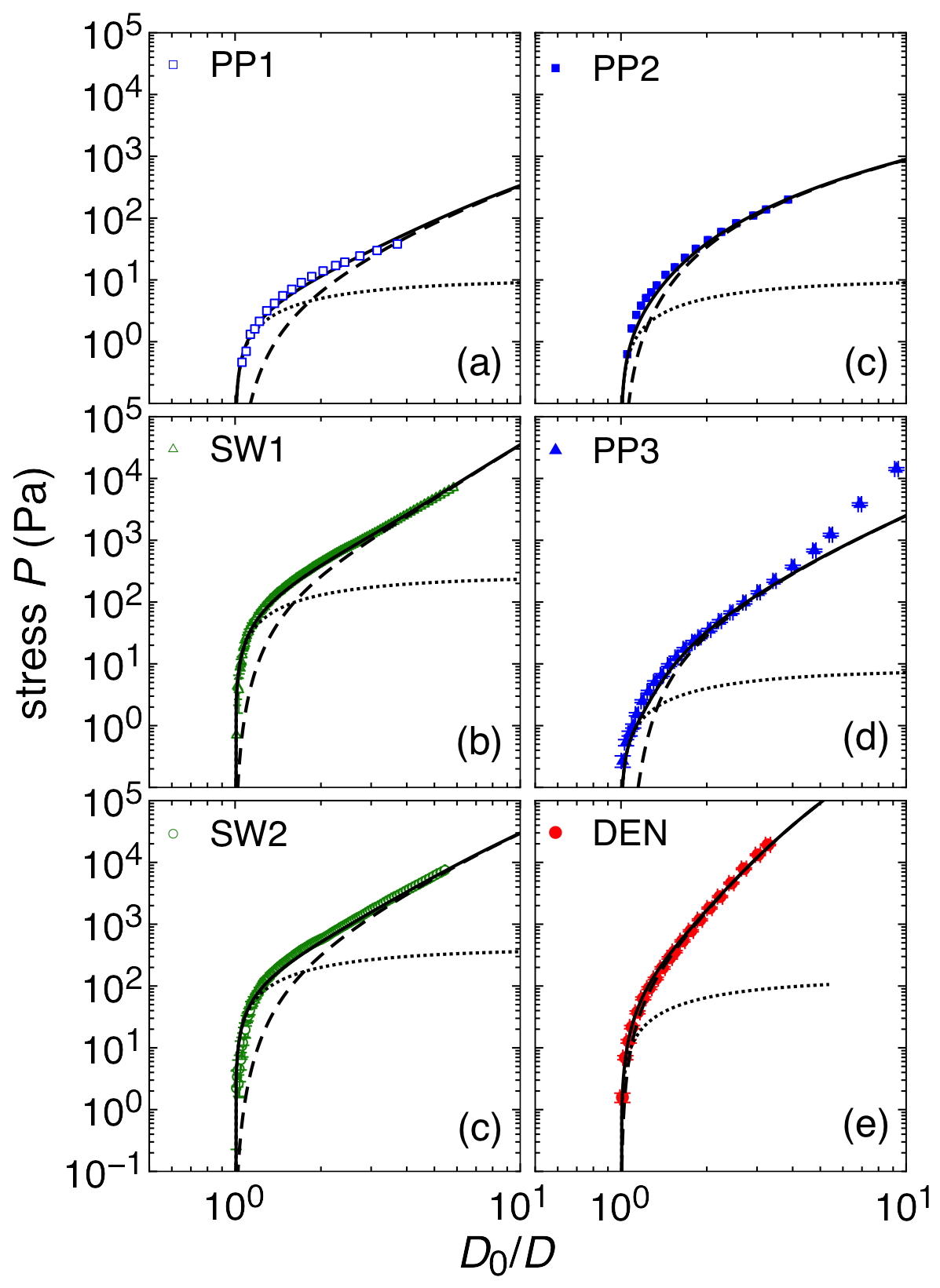}
\caption{Experimentally determined compression stress $P$ vs relative deformation $D_0/D$ for the six experimental fiber stacks studied in the paper: PP1 ($\textcolor{blue}{\square}$); PP2 ($\textcolor{blue}{\blacksquare}$); PP3 ($\textcolor{blue}{\blacktriangle}$); DEN ($\textcolor{red}{\bullet}$); SW1 ($\textcolor{ForestGreen}{\bigtriangleup}$); SW2 ($\textcolor{ForestGreen}{\bigcirc}$). Dashed lines represent the discrete self-consistent theory $P^{(3d)}_{\rm self}$ using experimentally determined spectrum, dotted line the linear elasticity behavior $P^{(3d)}_{\rm ref}$ and solid line the total stress $P_{\rm tot}(D)$.}
\label{figpexp}
\end{figure}

Stress-relative deformation curves are  given in Fig.~\ref{figdefphiexp} and Fig.~\ref{figpexp}. All materials exhibit a strong non linear elasticity, spreading over five decades of stress values. Using the experimentally determined spectra $\zeta_{0,q}$ and the value of $\kappa$, we apply the self-consistent model (see section \ref{selfconsistent}) to interpret the compression. The whole method is illustrated in Fig. \ref{figdefphiexp} in the case of the sample (SW1).

We assume that the stacks of fibers consist of the transverse sum of independent planes of effective density $\rho$, that can be estimated from the geometrical characteristics of the bundle (see Table \ref{stretch-table-1}). Since all planes contribute with $P^{\rm (2d)}(D)$ to the total stress $P^{\rm (3d)}(D)$, we write
\begin{equation}
P^{\rm (3d)}(D)= \rho P^{\rm (2d)}(D).
\end{equation}

For very weak relative deformations ($D_0/D\sim 1$), we observe a linear behavior corresponding to the deformation of the largest wavelength $\lambda_{\rm max}$. This linear elasticity is well described by $P^{\rm (3d)}_{\rm Ref} = \rho P^{\rm (2d)}_{\rm Ref}$, where $P^{\rm (2d)}_{\rm Ref}$ is given by eqn (\ref{pref}), and represented by dotted line in Fig. \ref{figdefphiexp} and \ref{figpexp}.

For larger relative deformations ($D_0/D \gg 1$), we test the predictive power of eqn (\ref{en}) by solving numerically the self-consistent eqn (\ref{sc}) for the experimentally determined distributions $\langle\zeta_{0,q}^2\rangle$. As a result we obtain the 2d-stress $P^{\rm (2d)}_{\rm self}(D)$ and $P^{\rm (3d)}_{\rm self}=\rho P^{\rm (2d)}_{\rm self}(D)$ that is represented as dotted-dashed lines in Fig. \ref{figdefphiexp} for SW1. If the overall shape of the data is well described by the theory, it is clear that it is necessary to introduce a scaling factor $\Phi_{\rm exp}$ to be able to reproduce the data for large deformation (see dashed line).

Finally, we calculate the total stress~:
\begin{equation}
P_{\rm tot}(D) = \rho\left(P^{\rm (2d)}_{\rm Ref}\left(D\right) + \Phi_{\rm exp} P^{\rm (2d)}_{\rm self}\left(D\right)\right).
\label{ptot}
\end{equation}
 where $\Phi_{\rm exp}$ is the only fitting parameters. $P_{\rm tot}(D)$ is represented as a solid line in Fig. \ref{figdefphiexp}. 

The experimentally determined compression stress $P$ and the theoretical analysis for the six experimental fiber stacks studied in the paper are given in Fig. \ref{figpexp} with same convention as for Fig. \ref{figdefphiexp}. For all studied samples, we observe a very good agreement between the three-dimensional experimental compression curves and the theoretical description.

\subsection{Numerical simulations}

To further understand how fiber shape disorder determines the compression of the stack, we performed numerical simulations on 2-dimensional systems. We investigated three classes of two-dimensional disordered systems. In this configuration, fiber rearrangements are forbidden. First, in order to validate our simulation method, we consider a stack of perfectly ordered sinusoidal fibers referred to as the {\sl reference system} (ref)  in the following (see Fig. \ref{figsimurefsmd} top inset).  As a second step, we introduce phase disorder to the system by choosing a random phase shift $\varphi_i$ homogeneously distributed in phase space $\varphi_i \in [0,2 \pi]$, referred to as {\it single mode disordered systems} (SMD).  Finally, we investigate fiber stacks with a {\sl power-law disorder} (PLD), inspired by the features of the experimentally measured distributions (Fig.~\ref{figspectre}). This allowed us, by varying the exponent $\alpha$ between 2 and 5, to test the self-consistent model by exploring a range of disorder wider than that of the experimental systems.

\subsubsection{Reference system (ref)}

\begin{figure}[h]
\begin{center}
\includegraphics[width=0.4\textwidth]{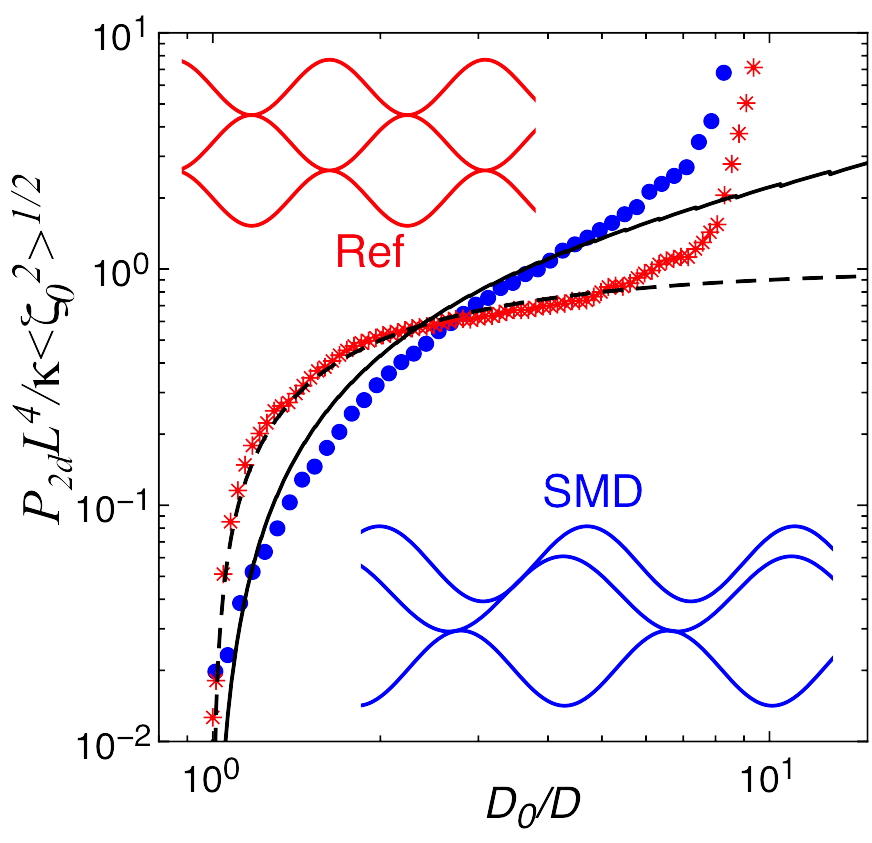}
\end{center}
\caption{(Color online) Numerical simulations for Ref and SMD and using $\kappa=2000\epsilon$: ($\textcolor{red}{\ast}$) normalized stress vs relative deformation $D_0/D$ for {\sl reference system} and (dashed line) linear theory $P_{ref}$ without any fitting parameter; ($\textcolor{blue}{\bullet}$) SMD systems with $\delta\lambda^2=0.01$ and (solid line) eqn (\ref{pressure_q_dis}) (see the ESI).}
\label{figsimurefsmd}
\end{figure}

Numerical simulation for the {\sl reference system} is shown in Fig.~\ref{figsimurefsmd} as ($\ast$) (see the ESI section S2 for more results). It exhibits linear elasticity, and the EOS is well described  in a large compression range,  without any fitting parameter, by eqn (\ref{pref}). At very high compressions, the fibers are fully squeezed and the stress is dominated by local excluded volume effects leading to a strong increase well above the bending contribution.

\subsubsection{Single mode disorder systems (SMD)}

Numerical simulations on disordered systems (SMD) distinctly exhibit non-linear elastic behavior (see Fig.~\ref{figsimurefsmd}). The compression behaviour is determined by randomness of phase shift, and the distribution of wavelengths plays only a minor role as shown in the ESI section S2. A simple analytical approach accounting for phase disorder for a sinusoidal spontaneous shape of wavelength $\lambda$ and amplitude $D_0$ allows calculating the normalized stress  as
\begin{equation}
\frac{P^{\rm (2d)}\lambda^4}{192\kappa D_0} = 2\int_{0}^{u_{max}} \frac{\left(1+\cos\left(2\pi u\right)-D/D_0\right)}{\left(1-2 u\right)^4 \left(1 +4 u\right)^2} \dd u
\label{pressure_q_dis}
\end{equation}
where $2\pi u_{max}=\arccos{\left(D/D_0-1\right)}$. 
Numerical simulation data is well described by eqn (\ref{pressure_q_dis}) without any fitting parameters. We also show in the ESI (section S2) numerical results for different fiber systems covering a range of parameters $\kappa$, $\lambda$ and $D_0$. 

\subsubsection{Power-law disorder system (PLD): numerical simulations}

\begin{figure}[h]
\begin{center}
\includegraphics[width=0.4\textwidth]{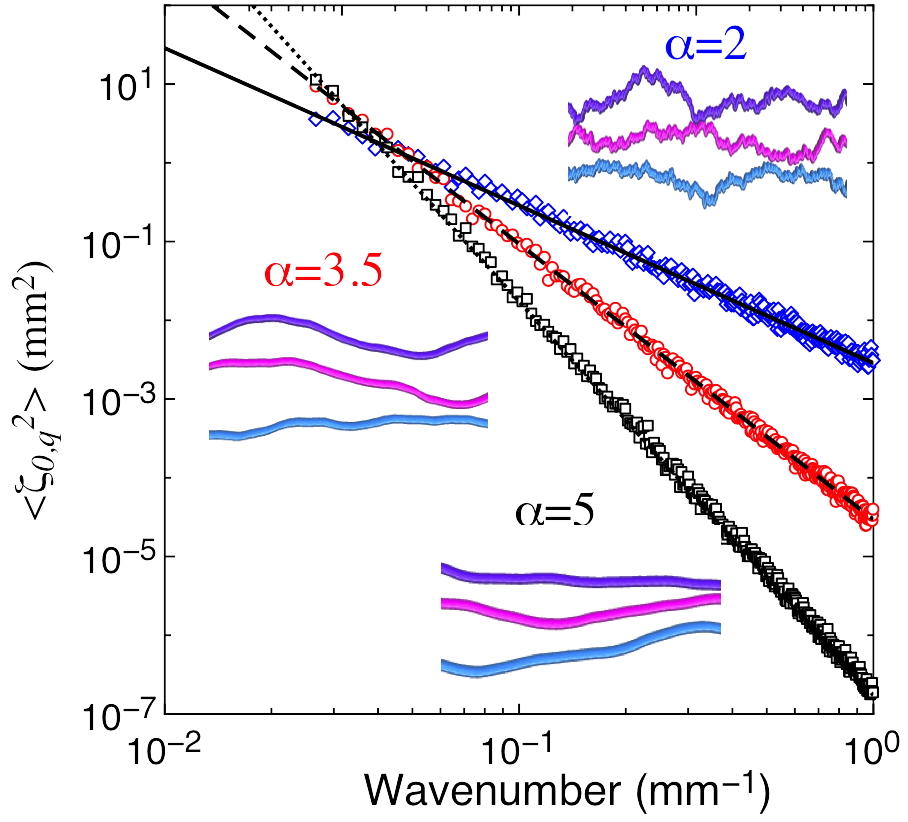}
\end{center}
\caption{(Color online) Amplitude $\zeta_{0,q}$ vs $q$ for PLD systems for $\alpha=2$ ($\textcolor{blue}{\lozenge}$), $\alpha=3.5$ ($\textcolor{red}{\circ}$) and $\alpha=5$ ($\textcolor{black}{\square}$). Also shown numerical typical fiber shapes.}
\label{figspectressimu}
\end{figure}

Finally, we investigate fiber stacks with a {\sl power-law disorder} (PLD), where the amplitude $\zeta_{0,q}$ of each mode follows a Gaussian probability distribution with mean square amplitudes
 \begin{eqnarray} 
\langle\zeta_{0,q_k}^2\rangle=\langle \zeta_0^2\rangle\left(\frac{\pi}{Lq_k}\right)^\alpha \zeta\left(\alpha,k_{min}+3/2\right)^{-1},
\label{zeta0theo}
\end{eqnarray}
with $q_k=\{q_{k_{min}},...q_{k_{Max}}\}$ ($k=\{k_{min},...k_{Max}\}$) and where $\zeta\left(\alpha,k_{min}+3/2\right)$ is the generalised zeta function \cite{1970_stegun_abramowitz}. Spectra are shown for different values of $\alpha$ in Fig. \ref{figspectressimu}. To avoid perfect stacking of fibers, we removed the first long wave-length mode $q<q_{k_{min}}$.

\begin{figure}[h]
\begin{center}
\includegraphics[width=0.4\textwidth]{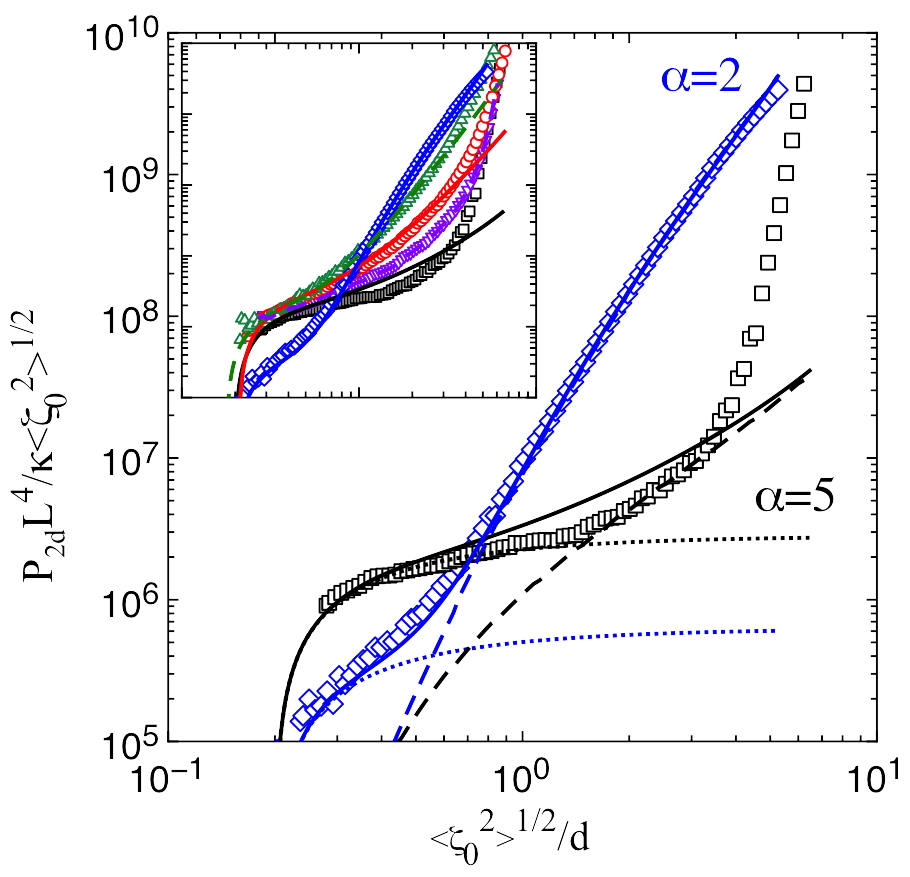}
\end{center}
\caption{(Color online) Normalized stress vs $\langle\zeta_0^2\rangle^{1/2}/d$ from numerical simulations for PLD systems for $\alpha=2$ ($\textcolor{blue}{\lozenge}$) and $\alpha=5$ ($\textcolor{black}{\square}$) and using $\kappa=2000\epsilon$. Dashed lines represent the discrete self-consistent theory $P_{\rm self}$ using numerical spectrum, dotted line the linear elasticity behavior $P_{\rm ref}$ and solid line the total stress $P_{\rm tot}=P_{\rm ref}$.  In inset same figure where lines represent the total stress obtained by adding the discrete self-consistent theory using numerical spectrum and the linear elasticity $P_{\rm ref}$ for all $\alpha$.}
\label{figstresssimu}
\end{figure}

Fig.~\ref{figstresssimu} displays compression results for the PLD cases in the normalized stress units $P_{\rm 2d} L^4/\kappa\langle\zeta_0^2\rangle^{1/2}$ as a function of the normalized density $\langle\zeta_0^2\rangle^{1/2}/D$, for $\alpha=2$ and $5$ (see also ESI section S2 for a more extensive set of $\alpha$ values). For the large density limit, where fibers are in close contact and where shape disorder is irrelevant, the data collapses on the same master curve, similarly to those of a single mode fiber -- see Fig.~\ref{figsimurefsmd}. For the most significant compression regime, at intermediate densities, we observe a strong dependence of the compression law on the value of the exponent $\alpha$, further confirming that the mechanical properties of the macroscopic stacks are controlled by fiber disorder.  

By following a procedure similar to that applied to fit the experimental results, we solve numerically the self-consistent relation eqn (\ref{sc}) for the numerical distributions $\langle\zeta_{0,q}^2\rangle$. Theoretical self-consistent results, presented in Fig.~\ref{figstresssimu} show a remarkable agreement with compression values from numerical simulations. The low compression regime is again well described by the linear elasticity $P_{\rm ref}^{\rm (2d)}$ without any fitting parameters. The self-consistent theory is in good agreement with the numerical simulations, especially for very rough fibers ($\alpha < 4$), provided that a global multiplication factor $\Phi_{\rm num}$ be applied as for experimental results. 

\section{Discussion}

\begin{figure}
\begin{center}
\includegraphics[width=0.4\textwidth]{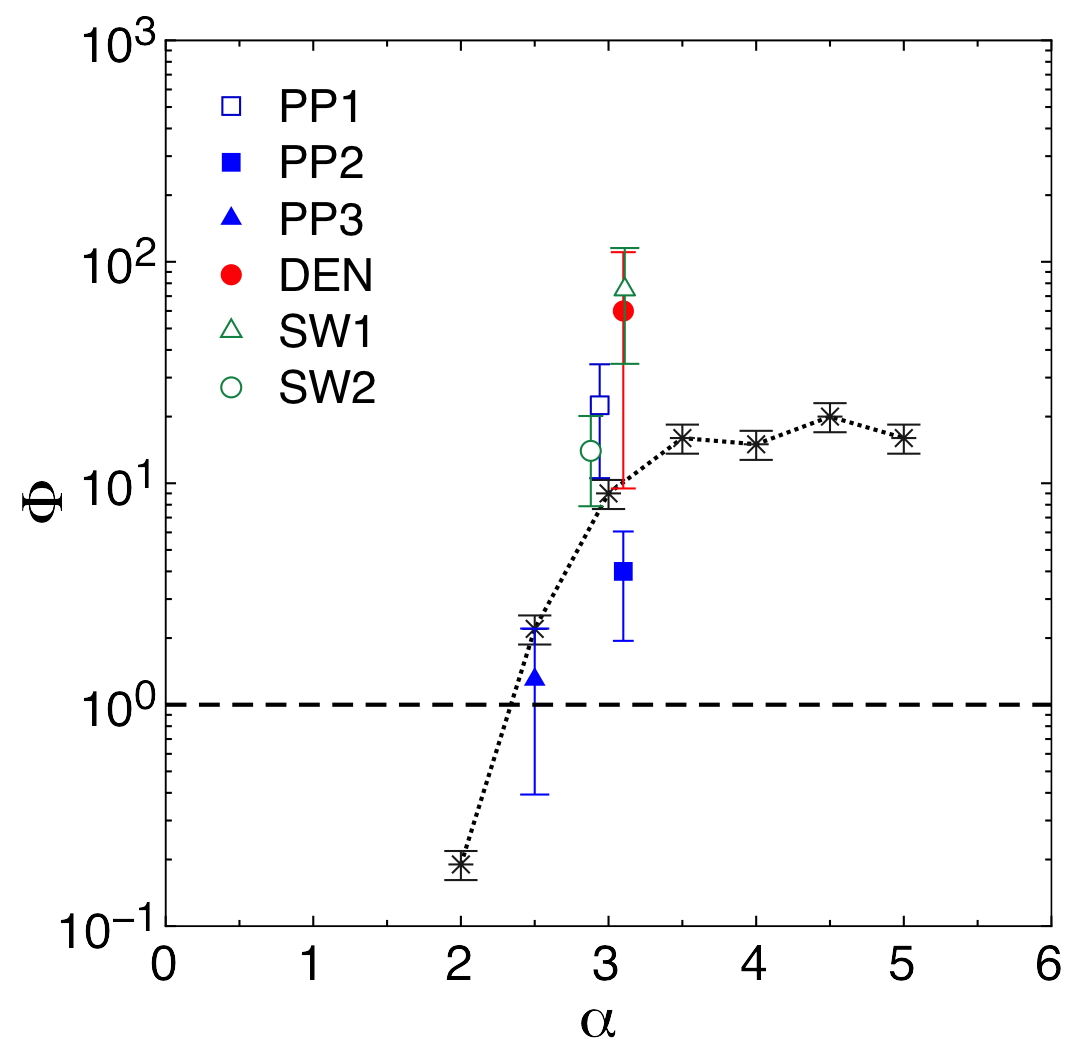}
\end{center}
\caption{(Color online) Normalization factors from the numerical simulations $\Phi_{\rm num}$ ($\ast$) and the experimental studies $\Phi_{\rm exp}$: PP1 ($\textcolor{blue}{\square}$); PP2 ($\textcolor{blue}{\blacksquare}$); PP3 ($\textcolor{blue}{\blacktriangle}$); DEN ($\textcolor{red}{\bullet}$); SW1 ($\textcolor{ForestGreen}{\bigtriangleup}$); SW2 ($\textcolor{ForestGreen}{\bigcirc}$).}
\label{figphinumphiexp}
\end{figure}

Both for experimental and numerical studies, systematic deviations between the mean-field predictions and the simulations can be seen in the low relative deformation limit, where the distance between fibers is of the order of the fibers mean-square amplitude. Mean-field theory poorly describes this limit, because of the vanishing number of fibers of mean amplitude larger than $\langle\zeta_0^2\rangle^{1/2}$. This regime can be qualitatively understood by noticing that, as the force rises sharply from zero, due to fiber-fiber contacts, it progressively builds up with essentially single-mode compression behavior. This is shown in Fig.~\ref{figpexp} and \ref{figstresssimu}, where the dashed line corresponds to the single mode expression $P_{\rm ref}$ with the wavelength $\lambda$ associated to the first mode of the distribution.

At larger relative deformations, in both experimental and numerical cases, the self-consistent theory correctly described our results up to a numerical scaling factor ($\Phi_{\rm exp}$ and $\Phi_{\rm num}$ respectively). In both cases, these are the unique fitting parameters that we introduce in the description. 

Fig.~\ref{figphinumphiexp} displays $\Phi_{\rm num}$ and $\Phi_{\rm exp}$ as a function of $\alpha$. Values for both coefficients are comparable to the experimental uncertainties. This clearly demonstrates that this coefficient is not related to experimental artifacts, further supporting that our 2 dimensional approach is also relevant to describe 3-dimensional experiments up to a consistent density of stress planes $\rho$, and that it is the self-consistent approximation that is involved. It is worth noting that the agreement between numerical simulations and self-consistent theory is optimal for $\alpha\sim2$, where $\Phi_{\rm num}$ is of order unit. This is further supported by the analytical resolution of the self consistent theory in continuous limit (see ESI, section S3). This corresponds to the case of the rougher fibers with the highest number of contacts, a situation where the mean-field self-consistent approximation is expected to be more accurate. The experimental fibers in this work have $\alpha$ values close to 2.8-3, were $\Phi$ is quite large. In future work, it will be interesting to identify classes of experimental fibers with $\alpha\sim 2$. It is also important to note that in our theoretical analysis and computer simulations all the fibers interactions are purely repulsive. In practice, environmental conditions such as humidity might add short range adhesive components (for instance capillary bridges) to the force between fibers. Without including drastic changes such as that from the long-range interactions in needled non-woven materials \cite{soares(2017)}, these short-range effects are nevertheless worth investigating in the future for their relevance for the compression of experimental systems such as lacquered hair with cross-linked contact points. Note that short-range attractions would also provide for material cohesion of the stacks, enabling for instance to perform tensile transversal experiments, 
an experimental geometry  better described by approaches for nonwoven fibrous mats\cite{siberstein(2012)}.

\section{Conclusions}

In summary we have investigated experimentally the mechanical properties of well aligned corrugated fiber stacks, showing that such fiber stacks display a strongly non-linear elastic behavior over 5-decades of stress. We showed that a theoretical self-consistent description \cite{2003_beckrich_charitat} connecting fiber shape disorder with stack compressibility explains well three different classes of fibers. We also performed numerical simulation studies for a larger class of fiber disorder. Interestingly we found that fiber shape, as characterized by the phase and wavelength disorder at fixed amplitude, is enough to induce a non-trivial compression behavior. We also simulated more realistic distributions for fiber disorder, with the spontaneous fiber shapes reconstituted from a superimposition of modes with power-law q-dependent amplitudes. For fibers of moderate corrugation the compression forces compare very well with our 2-dimensional mean-field theory. While more extensive simulation on 3-dimensional systems will certainly allow to better probe Van Wyk's seminal intuition \cite{1946_van-wyk}, our results here show that bending disorder indeed control the compressive behavior of aligned fibrous matter provided that the statistical nature of the fiber shape disorder is accounted for.

\section{Acknowledgements}

We wish to thank J\'er\'emie Geoffre and Damien Favier for their participation to the experiments. We gratefully acknowledge Joachim Wittmer, Jean Farago and Wiebke Drenckhan for fruitful discussions. N. S. thanks the Region Alsace for a PhD grant.
\bibliography{bib_fibers_cm.bib} 
\bibliographystyle{unsrt} 

\newpage



\section*{Supplementary material}

# Electronic supplementary information for The role of shape disorder in the collective behaviour of aligned fibrous matter

S. Salamone[1], N. Schulmann[1], O. Benzerara[1], H. Meyer[1], T. Charitat[11], C. M. Marques[1]

[1]UPR 22/CNRS, Institut Charles Sadron, Université de Strasbourg,
23 rue du Loess, BP 84047 67034 Strasbourg Cedex 2, France

## Contents

## S1 Single fiber characterisation

### S1.1 Eigenfunctions of the square Laplacian operator

In what follow, It is convenient to decompose fiber shape in the basis of the eigenfunctions $\Phi_{q_i}(x)$ of the square Laplacian operator $((\partial^2/\partial x^4 - q^4)\Phi_q(x) = 0)$ associated to curvature elasticity [1, 2].

$$\zeta_0(x) = \sum_i \zeta_{0,q_i} \Phi_{q_i}(x). \tag{S1}$$

Choosing boundary conditions ensuring that there are no forces nor torques on the strands we have

$$\Phi_{q_i}(x) = \frac{\cosh(\alpha_i) - \cos(\alpha_i)}{\sinh(\alpha_i) - \sin(\alpha_i)} \left[\sinh(q_i x) + \sin(q_i x)\right] - \left[\cosh(q_i x) + \cos(q_i x)\right], \tag{S2}$$

where the numerical coefficients $q_i L_0$ are determined from the relation $\cos(q_i L_0) \cosh(q_i L_0) = 1$. The solutions obey approximately $q_i = (i + 1/2)\pi/Na$ where $i \in \mathbb{N}^\star$. The wave vectors $q_i$ were picked uniformly with $i = i_{min}...N$.

### S1.2 Experimental fibers

#### S1.2.1 Individual fiber properties

An example of each type of fiber is given in Fig. S1. PP1 fibers appear to be highly disordered while PP2 fibers exhibit a dominant mode of wavelength 5 mm together with a distribution of wavelengths of smaller amplitudes. DEN fibers are more or less crenellated.

Transversal section diameters are of the order of 20 - 100 $\mu$m for PP fibers, 500 $\mu$m for DEN fibers and 200-300 $\mu$m for SW1 and SW2.

#### S1.2.2 Individual fiber shapes

Fiber shapes are characterised as follow. We first performed images of many fibers ($\sim 50$ per samples) using a scaner. We then extract the coordinate of fiber images to obtain the function $\zeta_0(x_i)$ that we expand numerically on the $\Phi_{q_i}(x)$ basis as eqn S1.

To check our method we have performed a test experiment as follow. We first impose an initial normal distribution of modes of variance $\zeta_{0,q}^{(ini)}$ given by

$$\langle \zeta_{0,q}^2 \rangle = \frac{c(\alpha)}{q^\alpha}, \tag{S3}$$

and corresponding to a Power Law Disorder (PLD). We choose randomly amplitudes $\zeta_{0,q}^{ini}$ following this normal distribution in order to plot a test fiber $\zeta_0^{(ini)}(x)$ (see Fig. ). We then printed and scanned this image as for experimental case (see Fig. S2(a)). We use the ImageJ software to extract the coordinates of the points which allows us to get $\zeta_0^{(num)}(x)$ (see Fig. S2). We then apply our transform analysis procedure and obtain the measured mode distribution $\zeta_{0,q}^{(mes)}$ (Fig. S2(d)):

$$\zeta_{0,q}^{(ini)} \longrightarrow \zeta_0^{(ini)}(x) \longrightarrow \zeta_0^{(num)}(x) \longrightarrow \zeta_{0,q}^{(mes)}. \tag{S4}$$

Results are plotted on Fig. S2(d), where we compare the initial mode distribution $\zeta_{0,q}^{ini}$, the calculated one $\zeta_{0,q}^{(mes)}$ and the initial power law distribution. It clearly demonstrate the validity of our method.

We present in Fig. S3(a) and (b) an example of shape analysis for PP1 fibers. Fig. S3(a) shows the spectra obtained for 4 different fibers. We also verify by simple visual observation that all the fibers of a stack are similar. Fig. S3(b) shows the comparison between the spectrum obtained for a fiber and that obtained by averaging over 40 fibers.

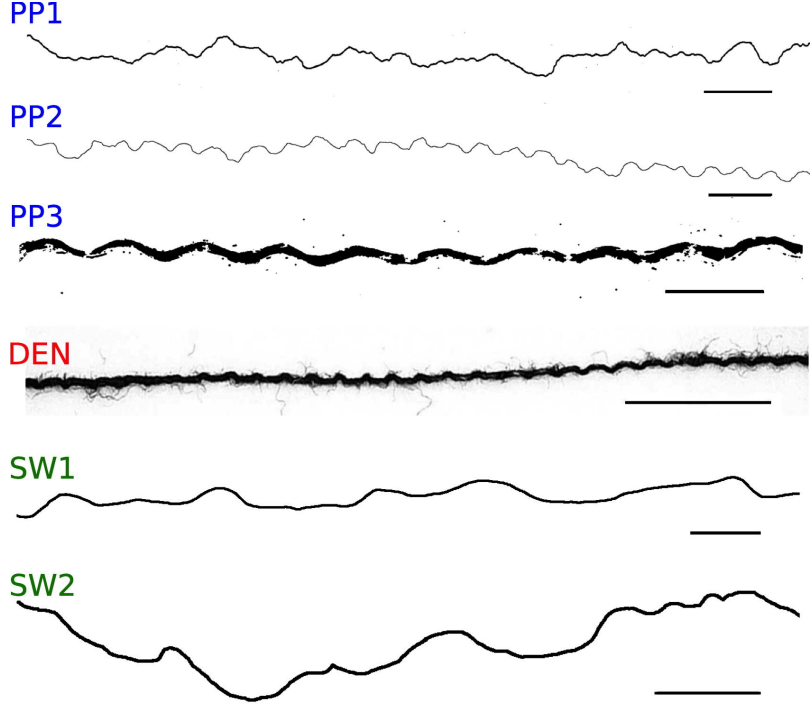

Figure S1: Individual images of all types of fibers studied: from top to bottom PP1, PP2, PP3, DEN, SW1 and SW2. Bar corresponds to 1cm.

#### S1.2.3 Videos of individual fiber bending modulus measurements

As described in the main paper, fiber bending modulus was measured by measuring oscillations of single fiber for SW1 and SW2 samples. A video showing an example of these experiments is shown as ESI (oscillation-FE1(fps25).avi).

For PP and DEN samples, we have developped a stretching method also described in paper. A video showing an example of these experiments is shown as ESI (etirement(fps25).avi).

### S1.3 Numerical fibers

We build the shapes of the isolated fibers from a given mathematical function $\zeta_0(x)$ along which we place $N$ monomers sequentially, at the equilibrium distance, $a = 1$, of the monomer-monomer bond interactions, setting at the same time the corresponding equilibrium angles $\theta_{0,i}$. At the end of this process, we have a fiber of $N$ monomers in an equilibrium shape close to the prescribed function $\zeta_0(x)$. In the following, we will consider three different cases for $\zeta_0(x)$. $L_0 = Na$ is the real length of the fiber, and $L$ its projection on the x-axis.

#### S1.3.1 Perfectly ordered Reference state (Ref)

We first consider fiber stacks of sinusoidal strands:

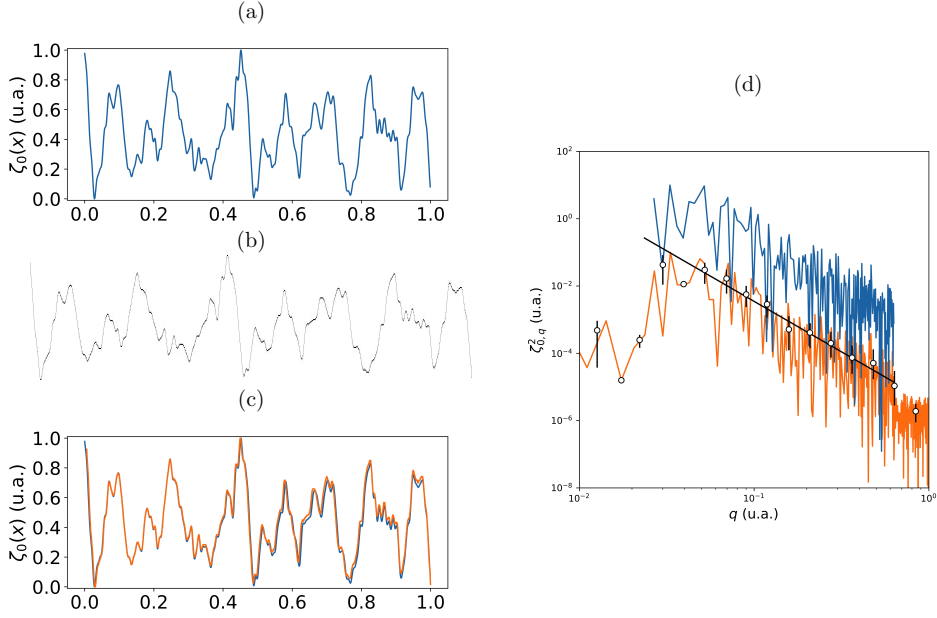

Figure S2: (a) Initial fiber $\zeta_0^{(\mathrm{ini})}(x)$ generated from a random spectra $\zeta_{0,q_i}^{(\mathrm{ini})}$. (b) Printed and digitized fiber $\zeta_0^{(\mathrm{num})}(x)$. (c) Comparison between (a) and (b). (d) Initial spectra $\zeta_{0,q_i}^{(\mathrm{ini})}$ (———) and numerical spectra $\zeta_{0,q}^{(\mathrm{mes})}$ obtained using our method. The amplitudes are shifted for sake of clarity. (∘) corresponds to moving average and (———) to initial power law.

$$\zeta_0^{(i)}(x) = d_0/2 \cos(2\pi x/\lambda_i + \varphi_i) \tag{S5}$$

where $\lambda_i$ and $\varphi_i$ are respectively the i$^{th}$ fiber's wavelength and the relative phase position.

In order to calibrate our simulation method against predictions from continuous elasticity theory, we test simple systems corresponding to single fibers with sinusoidal shapes, confined between parallel walls. In practice, for consistency with the more complicated systems where periodic boundary conditions are imposed, instead of simulating single chains between walls we simulate periodic systems consisting of pairs of chains with identical sinusoidal shapes and opposite phases

$$\forall i, \ \begin{cases} \lambda_i = \lambda \\ \varphi_i = (1-(-1)^i)\frac{\pi}{2} \end{cases}. \tag{S6}$$

The opposite phases coupled with chains small roughness due to the shapes of individual monomers insure that chains do not slide. In the following we will refer to this system as the *reference state* (Ref).

### S1.3.2 Disordered states

**i. Single mode disordered systems (SMD)** We consider a first class of disorder by adding a random translation phase shift $\varphi_i$ homogeneously distributed in phase space $\varphi_i \in [0, 2\pi]$ in equation S5. A small dispersion of wavelengths is also introduced ($\lambda_i$ are chosen from a Gaussian distribution of mean value

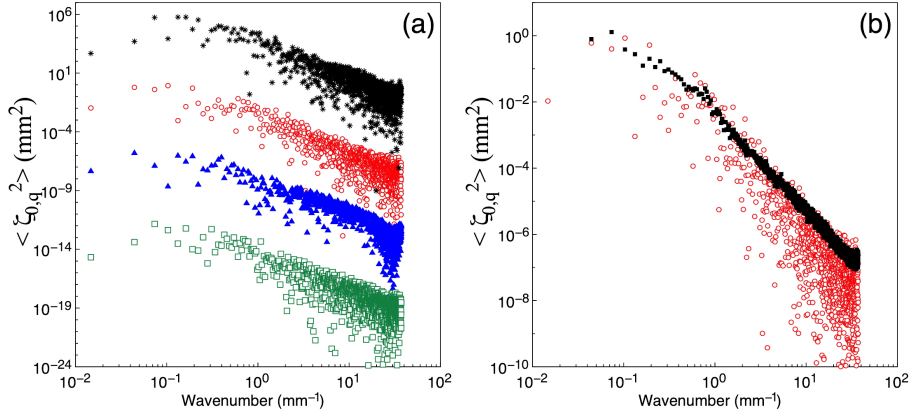

Figure S3: Shape analysis for PP1 samples: (a) $\zeta_{0,q}^2$ for 4 different PP1 fibers and (b) average spectra $\langle \zeta_{0,q}^2 \rangle$ over 40 fibers.

$\lambda_0$ and variance $\delta\lambda$) to avoid perfect stacking of fibers :

$$\varphi_i \in [0, 2\pi], \tag{S7}$$

$$\mathcal{P}(\lambda_i) = \frac{1}{\sqrt{2\pi\sigma^2}} e^{-\frac{(\lambda_i - \lambda_0)^2}{2\delta\lambda^2}}. \tag{S8}$$

**ii. Power-law disordered systems (PLD)** For *power-law disorder*, fiber shapes were obtained by constructing $\zeta_0(x)$ as a superposition of the eigenfunctions with amplitude $\zeta_{0,q_i}$ :

$$\langle \zeta_{0,q_i}^2 \rangle = \langle \zeta_0^2 \rangle \left(\frac{\pi}{Lq_i}\right)^\alpha \zeta(\alpha, i_{min} + 3/2)^{-1}, \tag{S9}$$

where $\zeta(\alpha, i_{min} + 3/2)$ is the generalized zeta function [3]:

$$\zeta(\alpha, a) = \sum_{i=0}^{+\infty} \frac{1}{(a+i)^\alpha}. \tag{S10}$$

We removed the largest wavelength mode $q \leq q_{i_{min}}$ to avoid fiber regular packing (we test $i_{min} = 1$ to 7) and choose to focus on systems with power-law distributions of wavelengths with associated amplitudes

We use values of the root-mean-square amplitude $\zeta_0 = \sqrt{\langle \zeta_0^2 \rangle}$ of a fiber in the continuous distribution model as a way of monitoring the vertical average extension of the fibers. $\langle \zeta_0^2 \rangle$ is given by :

$$\langle \zeta_0^2 \rangle = \frac{1}{L} \int_0^L dx \, \zeta_0(x)^2, \tag{S11}$$

where $L$ is the projected length of the fiber.

In order to check the consistency between analytical and numerical descriptions, a family of shapes of the type $\langle \zeta_{0,q_i}^2 \rangle = C/q_i^\alpha$ was generated and then the relation

$$\frac{\langle \zeta_0^2 \rangle}{C} = \zeta(\alpha, i_{min} + 3/2) \left(\frac{L}{\pi}\right)^\alpha. \tag{S12}$$

was plotted, see Fig. S4, with $\langle \zeta_0^2 \rangle$ measured by the smallest eigenvalue of the fiber gyration tensor.

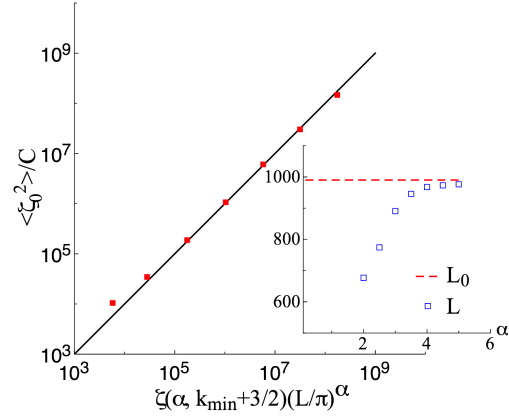

Figure S4: Normalized mean-square amplitudes $\langle \zeta_0^2 \rangle$ as a function of the disorder exponent $\alpha$ and comparison with the analytical expression S12. In the inset we compare the real chain length $L_0$ to its projection $L$ on the x-axis as a function of $\alpha$.

## S2 Supplementary numerical data

### S2.1 Reference state

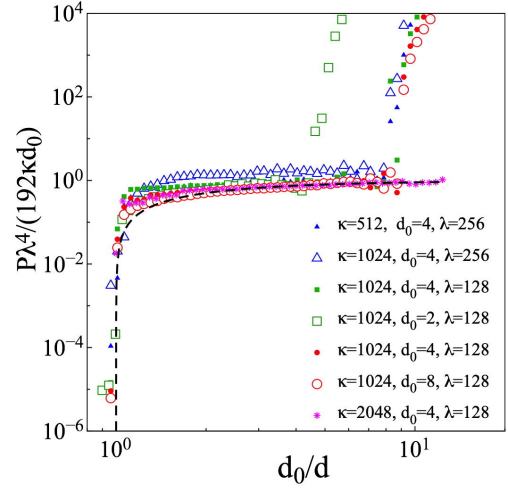

Figure S5: Normalized pressure $P\lambda^4/(192\kappa d_0)$ vs normalized density $d_0/d$ for monomode fiber stacks for various conditions of amplitude $\zeta_0$, bending modulus $\kappa$ and wavelength $\lambda$. The theoretical function $1 - d/d_0$ is also plotted for comparison.

## S2.2 Single mode disordered systems

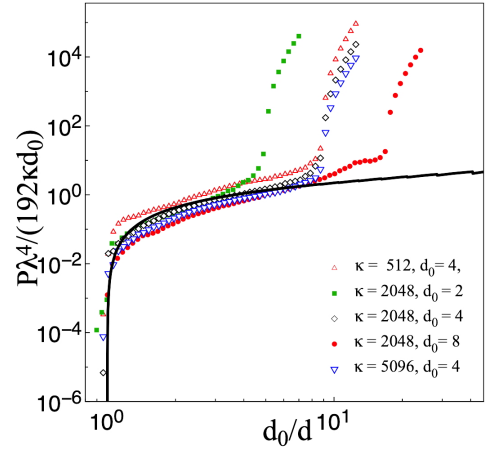

Figure S6: Normalized pressure $P\lambda^4/(192\kappa d_0)$ vs normalized density $d_0/d$ for single mode disordered systems for various conditions of amplitude $d_0$, bending modulus $\kappa$ and wavelength $\lambda$. The theoretical expression (equation [5] in main text) is also plotted for comparison.

## S2.3 Power-law disordered systems and comparison with discrete theory

Fig.S7 display complementary data for comparison between simulations and pressure obtained by solving numerically the self-consistent equation Eq. 2 (main text) for the experimentally determined distributions $\langle \zeta_{0,q}^2 \rangle$.

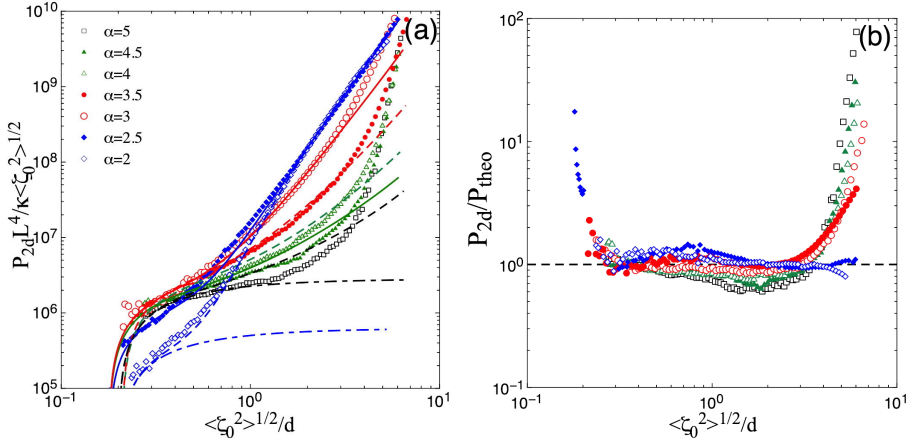

Figure S7: (a) Normalized pressure vs normalized density $\langle \zeta_0^2 \rangle^{1/2}/d$, comparison between numerical simulation data for $\alpha = 2$ to 5 and the sum of the numerical solution of the discrete self-consistent equations (equations 2 and 3 in main text) and the single mode contribution $P_{Ref}$ with wavelength $\lambda$ corresponding to the first mode of the distribution (dashed-dotted lines $P_{Ref}$ for $\alpha = 2$ and 5). (b) Ratio between numerical simulation pressure and discrete numerical solution of the self-consistent equation. Dashed line corresponds to the scaling solution.

## S3 Supplementary theoretical materials

### S3.1 Single-mode disorder model (SMD)

For this class of disorder we find numerically that even small perturbations in fiber ordering significantly modify the compression behavior of the fiber stacks in the intermediate compression regime as illustrated by data in Fig. S8.

Results for disordered systems in Fig. S8 suggest that the distribution of wavelengths plays only a minor role in the final compression behaviour that is instead determined by the presence of phase shift randomness. We propose here a simple analytical approach accounting for phase disorder by considering a test fiber with a sinusoidal spontaneous shape of wavelength $\lambda$ and amplitude $d_0$. Neighbouring fibers exerce localised forces on the test strand by direct contact, as sketched in Fig. (S8). We compute the deformation of the test fiber by minimizing the bending energy under a prescribed distance $d$ between fibers and a given phase shift measured by $x_0$, see Fig. S8. By defining the dimensionless variable $u = x_0/\lambda$ and integrating over all the relevant $x_0$ values we obtain the total pressure $P$ as:

$$\frac{P\lambda^4}{192\kappa d_0} = 2 \int_0^{u_{max}} \frac{(1 + \cos(2\pi u) - d/d_0)}{(1 - 2u)^4 (1 + 4u)^2} du, \tag{S13}$$

with $2\pi u_{max} = \arccos(d/d_0 - 1)$. Normalised numerical simulations data – see Fig. (S8) – follow a master curve well described by our prediction in Eq. (S13). We also obtained similar results for different fiber

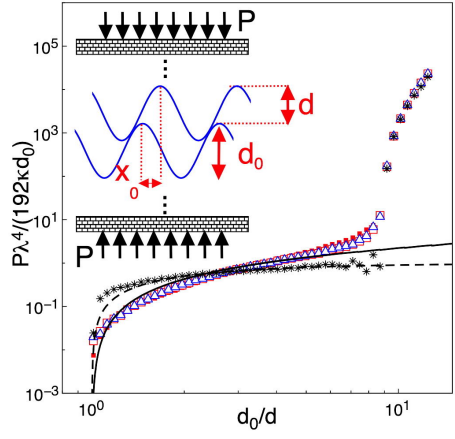

Figure S8: (Color online) Numerical simulation data for normalized pressure $P\lambda^4/(192\kappa d_0)$ vs normalized density $d_0/d$: ($*$) ordered systems of identical fibers and gaussian disordered systems with variances $\sigma = 0.001$ ($\square$), $\sigma = 0.01$ ($\blacksquare$) and $\sigma = 0.1$ ($\triangle$), see main text. All the simulations presented here have the same bending modulus $\kappa = 2048$, mean wavelength $\lambda = 128$ and amplitude $d_0 = 4$. The dashed line corresponds to the analytical expression $P_{Ref}$ (see main text) and the solid line to equation S13.

systems covering a range of parameters $\kappa$, $\lambda$ and $d_0$ (see Fig. S6).

### S3.2 Self-consistent equation for power-law disordered model (PLD)

#### S3.2.1 Scaling law analysis

For the disordered shapes described by Eq. S9, using for convenience dimensionless variables $x = d/\sqrt{\langle \zeta_0^2 \rangle}$, $\tilde{B} = BL^4/\kappa$ and $\tilde{P} = PL^4/\kappa\sqrt{\langle \zeta_0^2 \rangle}$ one gets

$$\langle e \rangle = g(\alpha) \frac{\kappa}{L^4} \frac{\langle \zeta_0^2 \rangle}{d} \left(\frac{BL^4}{\kappa}\right)^{(5-\alpha)/4}, \tag{S14}$$

$$\tilde{B}(x) = g(\alpha) \frac{\partial^2}{\partial x^2} \left(\tilde{B}(x)^{(5-\alpha)/4}\right), \tag{S15}$$

$$\tilde{P}(x) = -\int_{x_0}^{x} dx' \tilde{B}(x'), \tag{S16}$$

with

$$g(\alpha) = -\frac{\pi^{(2\alpha-3)/2}}{2^{(\alpha+1)/2}} \Gamma\left(\frac{7-\alpha}{4}\right) \Gamma\left(\frac{\alpha-5}{4}\right) \zeta\left(\alpha, \frac{5}{2}\right)^{-1}. \tag{S17}$$

Equations S15 and S16 have the scaling solution

$$\tilde{P}_\alpha(d) = c_p(\alpha) d^{-\frac{9-\alpha}{\alpha-1}}, \tag{S18}$$

with

$$c_p(\alpha) = \frac{(\alpha-1)}{(9-\alpha)} \left(\frac{2(9-\alpha)(5-\alpha)}{(1-\alpha)^2} g(\alpha)\right)^{4/(\alpha-1)}. \tag{S19}$$

Comparison between the scaling solution Eq. S18 and numerical simulations is shown on Fig. S11 for $\alpha = 2$ to $5$ as solid lines using an effective value $\alpha_{eff}$ as a fit parameter. Effective values $\alpha_{eff}$ are

plotted as a function of $\alpha$ on Fig. S12. This again demonstrates the good agreement between numerical simulations and self-consistent theory for $\alpha \sim 2-3$.

### S3.2.2 Complete resolution

The complete solution describing also the crossover to zero pressure can be obtained by numerically solving Eqs. S15 and S16. The compression modulus $B$ is given by:

$$B = \frac{\kappa}{L^4} \tilde{B}(x) = g(\alpha)^{4/(\alpha-1)} \frac{\kappa}{L^4} b(x), \tag{S20}$$

where the dimensionless modulus $b(x)$ is solution of the differential equation :

$$b(x) = \frac{\partial^2}{\partial x^2}\left[b(x)^{(5-\alpha)/4}\right]. \tag{S21}$$

As described in the main text, the scaling law $b(x) = b_0 x^{\beta-1}$ with

$$b_0(\alpha) = \left[2\frac{(9-\alpha)(5-\alpha)}{(1-\alpha)^2}\right]^{4/(\alpha-1)} \tag{S22}$$

$$\beta = \frac{9-\alpha}{1-\alpha}. \tag{S23}$$

is a particular solution of the differential Eq. S21.

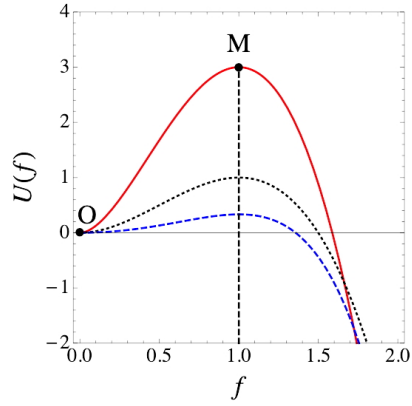

Figure S9: $U(f)$ vs $f$ for $\alpha = 2$ (red solid line), $\alpha = 3$ (black dotted line) and $\alpha = 4$ (blue dashed line).

To find a more general solution of Eq. S21, we define the function $g(x) = b(x)^{(5-\alpha)/4}$ which leads to :

$$g(x)'' = g(x)^{4/(5-\alpha)}. \tag{S24}$$

We look for a general solution of Eq. S24 as $g(x) = g_0 x^p f(x)$ where

$$g_0 = [p(\alpha)(p(\alpha)-1)]^{(5-\alpha)/(\alpha-1)}, \tag{S25}$$

with $p(\alpha) = 2(5-\alpha)/(1-\alpha)$. Assuming the variable change $x = e^{-t}$ we obtain the following differential equation for $f$ :

$$\ddot{f}(t) = -(1-2p)\dot{f}(t) + p(p-1)\left[f(t)^{1/\gamma} - f(t)\right], \tag{S26}$$

where $\dot{f}$ and $\ddot{f}$ are the first and second derivatives of $f$ with respect to $t$. This equation is analogous to the one dimensional motion equation of a particle located at $f$ in an external potential $U(f)$, with a fluid friction coefficient $(1-2p)$ (for $\alpha > 1/2$, $(1-2p) < 0$). The external potential $U(f)$ can be expressed as a function of $\alpha$ :

$$U(f) = 2\frac{(5-\alpha)(9-\alpha)}{(1-\alpha)^2}\left[\frac{1}{2}f^2 - \left(\frac{5-\alpha}{9-\alpha}\right)f^{(9-\alpha)/(5-\alpha)}\right], \tag{S27}$$

and has minimum for $f = 0$ and a maximum M at $f = 1$ (see Fig. S9).

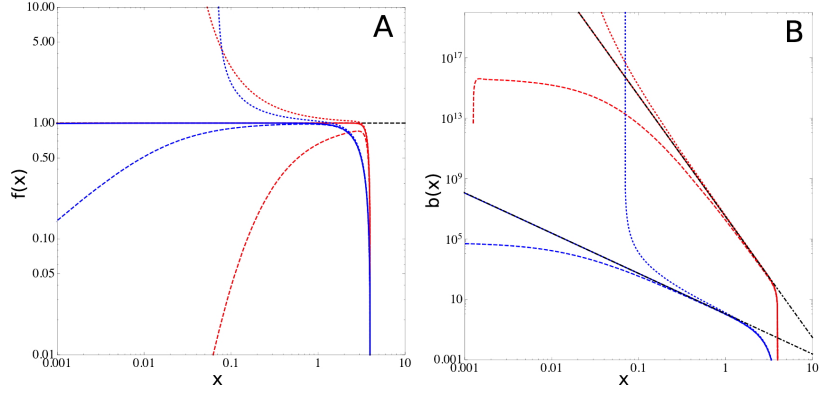

Figure S10: Numerical solution of differential equation S21. (a) $f(x)$ and (b) $b(x)$ for $\alpha = 2$ (red) and $\alpha = 4$ (blue). Solid line for $\dot{f}(t_0) = \dot{f}_c$, dotted line for $\dot{f}(t_0) > \dot{f}_c$ and dashed line for $\dot{f}(t_0) < \dot{f}_c$.

It is now possible to find a solution with a vanishing compression modulus $b(x_c) = 0$ ($f(t_0) = 0$) for $x_c = e^{-t_0}$ (generally $x_c \geq 1 \Leftrightarrow t_0 < 0$). The behavior of $f(t)$ depends on the second boundary condition. We discuss it now in terms of $\dot{f}(t_c)$. We have a particle leaving the point O ($f(t_0) = 0$) (compression modulus $b(x = x_c) = 0$) with an initial velocity $\dot{f}(t_0)$ :

- For $\dot{f}(t_0) = \dot{f}_c$ the initial *velocity* of the particle is sufficient to reach the maximum value M for $t \to +\infty$ ($x \to 0$). The particle tends to M and the scaling law $b_0 x^{\beta-1}$ is the asymptotic behavior of the compressibility (see Fig. S10.(a) and S10.(b), solid lines).

- For $\dot{f}(t_0) < \dot{f}_c$ the initial *velocity* of the particle is not large enough to escape from the minimum 0 and to cross the maximum value M. The particle goes close to M and then back to O. The compression modulus is non-monotonic and $b(x) \to 0$ for $x \to 0$ (see Fig. S10.(a) and S10.(b), dashed lines).

- For $\dot{f}(t_0) > \dot{f}_c$ the initial *velocity* of the particle is large enough to escape from the minimum 0 and to cross the maximum value M and the particle leaves the potential well. The compression modulus $b(x) \to +\infty$ for $x \to 0$ (see Fig. S10.(a) and S10.(b), dotted lines).

Figure S11 displays a direct comparison between this complete solution of the self-consistent equation and numerical simulations data, using as fitting parameters $x_c$ and $\alpha$. The best values of $\alpha$ denoted $\alpha_{eff}$ are represented as a function of $\alpha$ on Fig. S12 confirming again the good agreement between self-consistent theory and numerical simulations for $\alpha \sim 2-3$.

### S3.2.3 Comparison with numerical simulations

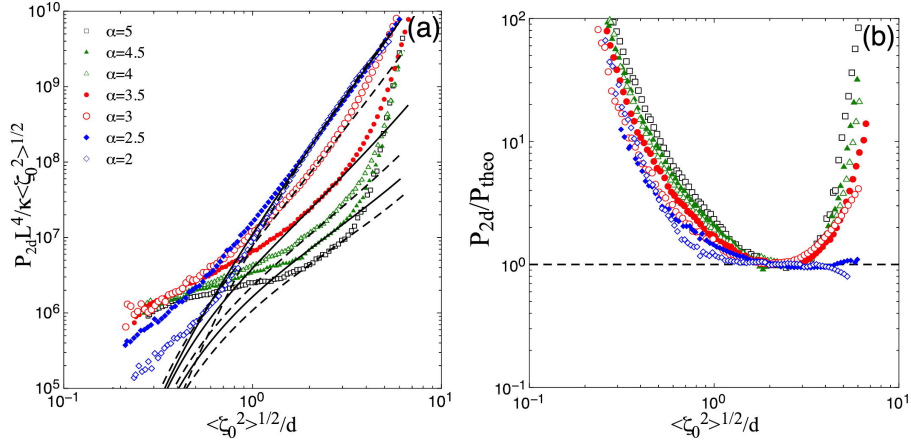

Figure S11: (a) Normalized pressure vs normalized density $\langle \zeta_0^2 \rangle^{1/2}/d$, comparison between numerical simulation data for $\alpha = 2$ to $5$ and solutions of continuous self-consistent theory with $\alpha_{\text{eff}}$ (see Fig. S12), (solid line) scaling law Eq.S18 and (dashed line) complete numerical solution of Eq. S21 using mathematical method described in section S3.2.2 (only for $\alpha = 2, 3, 4$ and $5$ for sake of clarity). (b) Numerical simulation pressure and numerical solution of self-consistent equation ratio. Dashed line corresponds to the scaling solution Eq.S18.

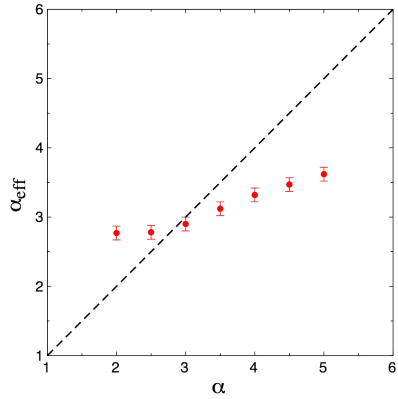

Figure S12: Effective exponent $\alpha_{\text{eff}}$ vs $\alpha$. Dashed line corresponds to $\alpha_{\text{eff}} = \alpha$.



\end{document}